\documentclass{article}
\usepackage{latexsym}
\usepackage{braket}
\usepackage{mathrsfs}
\usepackage{mathtools}
\usepackage{color}
\usepackage{xcolor}
\usepackage{bbm,bm}
\usepackage{amsmath}
\usepackage{amsfonts}
\usepackage{verbatim}
\usepackage{amssymb}
\usepackage{graphics,graphicx,xcolor}%
\usepackage{subcaption}
\usepackage{enumitem}
\usepackage[toc,page]{appendix}
\usepackage{cite}
\usepackage[english]{babel}
\usepackage{babelbib}
\usepackage{multirow}
\usepackage{tabularx}
\usepackage[T1]{fontenc}
\usepackage{soul}
\usepackage{makecell}
\setcellgapes{4pt}
\usepackage[a4paper, total={6in, 8in}]{geometry}

\usepackage[thinlines]{easytable}

\usepackage{colortbl}
\def\bal#1\eal{\begin{align}#1\end{align}}
\usepackage{color,soul}
\newcommand{\bsub}{\begin{subequations}}
\newcommand{\esub}{\end{subequations}}
\def\bal#1\eal{\begin{align}#1\end{align}}

 \newcommand{\grb}{{\beta}}  \newcommand{\grd}{{\delta}}
\newcommand{\gre}{{\epsilon}} \newcommand{\grz}{{\zeta}} \newcommand{\grh}{{\eta}} \newcommand{\gru}{{\theta}}
  \newcommand{\grl}{{\lambda}} \newcommand{\grm}{{\mu}}
\newcommand{\grn}{{\nu}} \newcommand{\grj}{{\xi}}  \newcommand{\grp}{{\pi}}
\newcommand{\grr}{{\rho}}  \newcommand{\grt}{{\tau}} 
\newcommand{\grf}{{\phi}}  \newcommand{\grc}{{\psi}} \newcommand{\grv}{{\omega}}
   \newcommand{\grD}{{\Delta}}
   
  \newcommand{\grL}{{\Lambda}}

  \newcommand{\grC}{{\Psi}} \newcommand{\grV}{{\Omega}}

\date{}

\begin{document}

\title{Time-covariant Schr\"{o}dinger equation and invariant decay probability: The $\Lambda$-Kantowski-Sachs universe}
\author{Theodoros Pailas $^{1}$, Nikolaos Dimakis $^{2}$, Petros A. Terzis $^{3}$, Theodosios Christodoulakis $^{4}$\\
\normalsize
{$^{1,3,4}$\it Nuclear and Particle Physics Section, Physics Department, University of Athens, 15771 Athens, Greece.}\\
{$^2$\it \normalsize{\it}{Center for Theoretical Physics, College of Physics, Sichuan University, Chengdu 610064, China.}}\\
\normalsize{\it }}

\maketitle

\abstract{The system under study is the $\Lambda$-Kantowski-Sachs universe. Its canonical quantization is provided based on a recently developed method: the singular minisuperspace Lagrangian describing the system, is reduced to a regular (by inserting into the dynamical equations the lapse dictated by the quadratic constraint) possessing an explicit (though arbitrary) time dependence; thus a time-covariant Schr\"{o}dinger equation arises. Additionally, an invariant (under transformations $t=f(\tilde{t})$) decay probability is defined and thus ``observers'' which correspond to different gauge choices obtain, by default, the same results. The time of decay for a Gaussian wave packet localized around the point $a=0$ (where $a$ the radial scale factor) is calculated to be of the order $\sim 10^{-42}-10^{-41}\mathrm{s}$. The acquired value is near the end of the Planck era (when comparing to a FLRW universe), during which the quantum effects are most prominent. Some of the results are compared to those obtained by following the well known canonical quantization of cosmological systems, i.e. the solutions of the Wheeler-DeWitt equation.}

\newpage

\section{Introduction}

\setstcolor{red}
\setul{}{2pt}

Through the years, numerous attempts have been made to quantize gravity. Some of them more conservative, \cite{10.2307/100496,Dirac:1958sc,PhysRev.116.1322,PhysRev.116.1324,PhysRev.116.1325,PhysRev.116.1326,PhysRev.116.1327,PhysRev.116.1328,PhysRev.116.1329,PhysRev.116.13210,Wheeler:1964qna, Wheeler:1988zr, DeWitt:1962cg, PhysRev.160.1113, PhysRev.162.1195, PhysRev.162.1239} while others more radical \cite{Schwarz_2007,Ib_ez_2000,green_schwarz_witten_2012,DUFF_1996,rovelli_2004,Thiemann_2007,Rovelli_1998,Loll_1998,Ashtekar2021}. More or less, all of them share a number of basic reasons to pursue such a quest. The more fundamental are: the expectation that at some energy scale, all the fundamental interactions should be unified. Thus, there must exist a quantum theory of gravity in the picture of quantum interactions. The second motivation is based on the existence of abnormalities in the classical General Relativity and/or its modifications. By abnormalities we refer to the singularities of black holes or the Big Bang itself for instance. The third and final is the incompatibility of General Relativity with Quantum Mechanics (and even more Quantum Field Theory) regarding the notions of space and time. In the former, spacetime is a dynamical entity that affects and, at the same time, is affected by any form of energy, while in the latter, it is an unaltered ``frozen'' and ``external'' arena. For some interesting reviews on these matters we recommend \cite{PhysRevD.40.2598,Kuchar:1991qf,Kiefer:2013jqa,Isham:1992ms,rovelli_2004,Giulia}. Also, for an interesting new approach regarding the notion of time and resolution of singularities one may look  at \cite{H_hn_2020, Hoehn:2019owq,Gielen} and references therein.

To tackle the problem by considering the full theory of Gravity is a noble,  but quite difficult task; thus, over the years, researchers have chosen to study simplified cases. The simplest appear to be the minisuperspace models, which are obtained via symmetry reduction of the full theory. By assuming a certain level of symmetry applied on the spacetime metric and the matter fields, the system's degrees of freedom are reduced to finite from infinite. Thus, instead of field theory, the tools of point-like mechanics can be used. Such kind of spacetimes are the FLRW geometry, the Schwarzschild and Reissner-Nordstr\"{o}m black holes and the anisotropic Bianchi and Kantowski-Sachs types \cite{Ryan,Schwarzschild:1916uq,1916AnP...355..106R}. There are many works dealing with such kind of simplified cases, with very interesting results. It is quite difficult to cite them all, so we just mention a handful of them: The quantization of the FLRW geometry in the presence of a scalar field has been studied extensively from various perspectives \cite{Hawking1984,Page:1990mh,Ashtekar2006,PintoNeto2007,Kiefer2012,Vakili2012,Kim2014,Gryb2019}. An interpretation via the de Broglie-Bohm theory of the solutions to the Wheeler-DeWitt equation for the Reissner-Nordstr\"{o}m-de Sitter black hole was presented in \cite{Kenmoku:1998ax}. The authors of \cite{Christodoulakis:1991jd,Christodoulakis:2001um,Christodoulakis:2013sya} exploit the notion of conditional symmetries to provide a quantum description of some Bianchi types and of the Reissner-Nordstr\"{o}m black hole. There were also attempts in providing a generalized definition of probability \cite{Dimakis:2016mpg} and also a quantum description through the definition of a time related to the homothetic symmetry of the minisuperspace metric \cite{Karagiorgos:2018gkn}. Quantization in a larger space through an Eisenhart lift was recently employed in \cite{Kan2021}. An analogy between the Schwarzschild- Reissner-Nordstr\"{o}m black holes and hydrogen atoms were drawn in \cite{Corda:2019vuk,Corda:2020fjz}, where their energy spectrum was obtained based on some constructed Schr\"{o}dinger type of equation. A study related to a mass operator and the existence of non-zero mass uncertainty can be found in \cite{Davidson:2014tda,Davidson:2012dt}. The quantization of an inhomogeneous cosmological model has been performed in \cite{Marugan1997,Hybrid2008,Szek_quant,Pal_Zamp_Sz}.

In this work we are dealing with a Kantowski-Sachs spacetime. Various aspects of the canonical quantization of Kantowski-Sachs models have been previously discussed in the literature: Louko and Vachaspati \cite{exte1} studied whether or not the Vilenkin boundary condition could provide a unique wave function for the Kantowski-Sachs minisuperspace model. In \cite{exte2} the authors provide a method for approximate evaluation of the path integral for spatially homogeneous minisuperspace models and apply it to the Kantowski-Sachs geometry among others. Finally, the complete solutions to the Wheeler-DeWitt equation were provided in \cite{exte3} for generalized Kantowski - Sachs models with cosmological constant and pressureless dust. Several works also exist in the quantization of cosmological models in extended theories of gravity \cite{Shojai2008,PaliathanasisBD,Xu2016,Darabi2019,Capozzielloft,Paliathanasisftb}.

As any gauge theory \cite{pokorski_2000, weinberg_1995, Maggiore_845116} contains spurious degrees of freedom, so does gravity. This results to its description via a singular Lagrangian, which is equivalent to the existence of constraint equations. The question then is how we deal with the constraints during the quantization procedure. There are two main approaches in this problem, which are nicely described in \cite{Isham:1992ms}: The first approach is called the reduced phase space quantization. This is based on trying to find some canonical (invertible) transformation from the initial canonical variables to a complete set of gauge invariant combinations, plus a set of specific functions of the spacetime coordinates. The latter implies that the gauge freedom is completely broken and the constraint equations are strongly solved (since the reduced Hamiltonian is expressed solely in terms of the dynamical degrees of freedom) with the introduction of the Dirac Brackets. Thus, on the embedded hypersurface defined by the gauge choices, a Schr\"{o}dinger equation can be constructed. This can be extended by considering the true degrees of freedom as functions of the spatial coordinates on the hypersurfaces, leading eventually to the so called ``multi-finger'' Schr\"{o}dinger equation \cite{Isham:1992ms, PhysRev.126.1864, Kuchar:1971xm}. Note that this generalization is possible only in the case of field theory and not for the minisuperspace models. In many cases it is quite difficult to obtain the reduced Lagrangian, thus many authors have turned to the definition of a time standard based on matter fields \cite{Brown:1994py, Alexander:2012tq, Bojowald:2010xp, Rovelli:1989jn,Kiefer:1988ud}. To the advantages of this procedure are included: the capability of defining a Hilbert space and thus provide a probabilistic interpretation, as well as the dependence of the Hamiltonian only on the true degrees of freedom. The main disadvantage is the existence of many different quantum descriptions for the different gauge choices, which are not equivalent by default \cite{Barvinsky:1993jf,Pons:2009cz}. This is related to the fact that there is no unique canonical transformation, which raises problems at the quantum level, due to the non-commutative nature of the order in which you perform the transformation and the quantization. The second approach is covariant, thus by definition it holds for every gauge choice. Being covariant implies that the constraints are carried to the quantum level and imposed to the physical states as restrictions, resulting eventually to the so called Wheeler-DeWitt equation \cite{10.2307/100496,Dirac:1958sc,Wheeler:1964qna, Wheeler:1988zr, DeWitt:1962cg, PhysRev.160.1113, PhysRev.162.1195, PhysRev.162.1239}. The main problems are the absence of a well defined Hilbert space and, since time is not an external parameter, the inability to obtain evolution of states in the usual quantum mechanical sense. For a discussion related to the differences between covariant and reduced phase space quantization see \cite{exte4,exte5}.

The present work constitutes an extension of a relatively recent published paper \cite{Pailas:2020msz} regarding the Reissner-Nordstr\"{o}m black hole. The main idea was to somehow render the constraint equation trivial, in the sense that it will be satisfied modulo the dynamical equations, while at the same time maintain intact the gauge invariance. The result was that the singular Lagrangian describing the Reissner-Nordstr\"{o}m black hole reduced to a regular one with an explicit ``time'' dependence (note that ``time'' for the Reissner-Nordstr\"{o}m case is actually the radial distance, hence the quotations). The parameter of ``time'' was identified by considering the remaining gauge degrees of freedom as mere functions of it and not as dynamical quantities. Based on this method, a ``time''-covariant  Schr\"{o}dinger equation was constructed, giving a somewhat intermediate picture in comparison to the two previous approaches. The covariance, in what regards purely ``time'' reparametrizations, is maintained, while a Schr\"{o}dinger equation incorporating each different gauge choice of this type is produced. Here, we are interested to apply this method to a cosmological minisuperspace model and specifically to the geometry of $\grL$-Kantowski-Sachs universe. Furthermore, we extend it a bit further by defining an invariant decay probability. In the end, we also present the \emph{de facto} covariant method of the typical Wheeler-DeWitt quantization of the given model, in order to study where the two approaches meet and what differences there exist. At this point, we need to mention of another interesting work, where a similar procedure with the one we follow here is employed. It regards the derivation of a Schr\"{o}dinger equation and the introduction of time dependence in the wave function through the process of gauge fixing the scale factor while leaving the lapse as a degree of freedom, for more details see \cite{Davidson3}.

The paper is organized as follows: In section 2 the classical description of the system is provided by deriving the solutions and the relative reduced Lagrangians and Hamiltonians. The canonical quantization procedure is presented in section 3 alongside with what we call the time-covariant Schr\"{o}dinger equation and its solutions.  Next, in section 4, the invariant probability decay rate is calculated for the case of a Gaussian initial state. Additionally, an estimate on the inflation epoch based on the quantum effects is presented. In section 5, we briefly revisit the typical Wheeler-DeWitt quantization for a comparison to the previous approach and in the last section we gather our conclusions.


\section{Classical description}

\subsection{Solutions}

Our starting point is the Einstein-Hilbert action in the presence of a cosmological constant $\Lambda$,
\begin{equation} \label{EHaction}
  S_{EH} = \int d^4x \sqrt{-g}\mathcal{L}_{EH} = \int d^4x \sqrt{-g}\frac{1}{2\kappa}\left(-R + 2 \Lambda \right),
\end{equation}
where $g=\det(g_{\mu\nu})$ is the determinant of the space-time metric and $R$ is the Ricci scalar. For simplicity, in what follows we adopt the units $\kappa=c=1$.

We consider the following line element, in spherical coordinates $x^{\grm}=(t,r,\gru,\grf)$,
\begin{align}
ds^{2}=g_{\mu\nu}dx^\mu dx^\nu=-n^{2}dt^{2}+a^{2}dr^{2}+b^{2}\left(d\gru^{2}+\sin^{2}\gru\,d\grf^{2}\right),\label{eq1}
\end{align}
where the degrees of freedom $(n,a,b)$ are functions of the coordinate $t$.

With this ansatz, Einstein's equations
\begin{equation}
  R_{\grm\grn}- \frac{1}{2} R g_{\mu\nu}+\grL g_{\grm\grn}=0,
\end{equation}
in which $R_{\mu\nu}$ represents the Ricci tensor, reduce to a set of ordinary differential equations for the functions of $t$ entering the metric. The field equations are equivalent to:
\begin{align}
&\frac{(1-\grL b^{2})n^{2}}{b^{2}}+\frac{2\dot{a}\dot{b}}{a\,b}+\frac{\dot{b}^{2}}{b^{2}}=0,\label{eq2}\\
&\grL a^{2}-\frac{a^{2}}{b^{2}}-\frac{a^{2}\dot{b}^{2}}{b^{2}n^{2}}+\frac{2a^{2}\dot{b}\dot{n}}{b n^{3}}-\frac{2a^{2}\ddot{b}}{b n^{2}}=0,\label{eq3}\\
&\grL b^{2}-\frac{b \dot{a}\dot{b}}{a n^{2}}+\frac{b^{2}\dot{a}\dot{n}}{a n^{3}}+\frac{b\dot{b}\dot{n}}{n^{3}}-\frac{b^{2}\ddot{a}}{a n^{2}}-\frac{b \ddot{b}}{n^{2}}=0.\label{eq4}
\end{align}
For our purposes, by following the same steps as in \cite{Pailas:2020msz}, the constraint equation \eqref{eq2} should be solved, if possible, with respect to the lapse $n$. It turns out that this proposal was not unknown and comes by the acronym BSW from the initials of Baierlein, Sharp and Wheeler \cite{PhysRev.126.1864}. A novel derivation of General Relativity, partially based on the above method can be found in \cite{Barbour_2002}. In our case, due to the presence of the cosmological constant, this is possible only for $b(t)\neq\pm\frac{1}{\sqrt{\grL}}$. From now on, without loss of generality, we consider $b(t)>0$ and consider the cases $b(t)=\frac{1}{\sqrt{\grL}}$ and $b(t)\neq \frac{1}{\sqrt{\grL}}$.

Before we proceed with the study of these branches, we note that the above set of equations can be derived via the variation of the following minisuperspace action
\begin{align}
&S_{min}=\int{L_{EH}(n,a,b) dt},\label{eq5}\\
&L_{EH}=\frac{1}{n}\left(2b\dot{a}\dot{b}+a \dot{b}^{2}\right)+n\left(\grL b^{2}-1\right) a.\label{eq6}
\end{align}

This action can be obtained from \eqref{EHaction} by dropping out the spatial, non-dynamical part of the integral, i.e.
\begin{equation} \label{fromEHtomin}
  S_{EH} = \int d^4x\sqrt{-g}\mathcal{L}_{EH} = \int \sin\theta dr d\theta d\phi \int L_{EH}(t) dt =\left(\int \sin\theta dr d\theta d\phi \right) S_{min}.
\end{equation}
The dynamics of the system is completely encoded in the reduced action $S_{min}$ describing a system of finite degrees of freedom. We now continue with the classical analysis of the solution space for this system.

\subsubsection{Case 1, $b(t)=\frac{1}{\sqrt{\grL}}$. Bertotti-Kasner space }

For this specific choice, the number of independent Einstein equations reduces to just one second order differential equation
\begin{align}
\ddot{a}-\dot{a}\frac{\dot{n}}{n}-\grL a n^{2}=0.\label{eq7}
\end{align}
Due to the transformation law of the lapse function and the $t$ derivatives, this equation is still covariant under transformations of the form $t=f(\grt).$ Regarding it's solution, it can be solved  with respect to either $n$ or $a$. However, if another degree of freedom $s$ is defined via $s=\int{n dt}$, the solution with respect to $a$ assumes the elegant form:
\begin{align}
&a(t)=c_{1}e^{\sqrt{\grL}s(t)}+c_{2}e^{-\sqrt{\grL}s(t)},\label{eq8}\\
&ds^{2}=-\dot{s}^{2}dt^{2}+\left(c_{1}e^{\sqrt{\grL}s(t)}+c_{2}e^{-\sqrt{\grL}s(t)}\right)^{2}dr^{2}+\frac{1}{\grL}\left(d\gru^{2}+\sin^{2}\gru d\grf^{2}\right),\label{eq8}
\end{align}
where $c_{1},c_{2}$ constitute non-essential (absorbable via coordinate transformations) constants. This solution is called the Bertotti-Kasner space \cite{1998PhLA..245..363R,PhysRev.116.1331} and belongs to the family of Kantowski-Sachs spacetimes \cite{Kantowski:1966te}.

\subsubsection{Case 2, $b(t)\neq\frac{1}{\sqrt{\grL}}.$ The generic solution}

Contrary to the previous case, we can now solve the constraint equation with respect to $n$:
\begin{align}
n=\frac{\sqrt{2b\dot{a}\dot{b}+a{\dot{b}}^{2}}}{\sqrt{a\left(\grL b^{2}-1\right)}}.\label{eq9}
\end{align}
When this value of $n$ is substituted into the rest of the equations, only one of the reduced equations is independent, and one way to express it is
\begin{align}
\ddot{a}=\dot{a}\frac{\ddot{b}}{\dot{b}}-\frac{{\dot{a}}^{2}}{a}+\frac{2\dot{a}\dot{b}}{b\left(\grL b^{2}-1\right)}+\frac{\grL a {\dot{b}}^{2}}{\grL b^{2}-1}.\label{eq10}
\end{align}
This equation is covariant as in the previous case. The above expression is simplified if we introduce the following degree of freedom $w$
\begin{align}
&a=\sqrt{\frac{w}{b}},\,w=a^{2}b,\label{eq12}
\end{align}
so that the previous equation becomes
\begin{align}
\ddot{w}=\frac{\ddot{b}}{\dot{b}}\dot{w}+\frac{2\grL b \dot{b}\dot{w}}{\grL b^{2}-1}.\label{eq13}
\end{align}
These variables have been used before in \cite{Alonso_Serrano_2014} for the study of correlations across the horizons that this solution possesses. In these variables, the solution of \eqref{eq13} is rather simple
\begin{align}
&w(t)=M-b(t)+\frac{\grL}{3}b(t)^{3},\label{eq14}\\
&ds^{2}=-\frac{\dot{b}(t)^{2}}{-1+\frac{M}{b(t)}+\frac{\grL}{3}b(t)^{2}}dt^{2}+\left(-1+\frac{M}{b(t)}+\frac{\grL}{3}b(t)^{2}\right)dr^{2}+b(t)^{2}\left(d\gru^{2}+\sin^{2}\gru d\grf^{2}\right).\label{eq15}
\end{align}
In the above line element, the constant $M$ is essential. Note that usually $M$ is associated to a ``mass'' constant $m_0=\frac{M}{2}$. Here for simplicity and in order to avoid more numeric factors we just use $M$ for the constant of integration.

\subsection{Lagrangian-Hamiltonian description}

The starting point to construct a Lagrangian in order to describe the dynamics of the previous system is the Einstein-Hilbert action. As we can see from \eqref{eq6}, such a Lagrangian exists and we have to find out its reduced form after the assumptions of Case 1 and the solution of the constraint in Case 2.

\subsubsection{Case 1}

There are two basic ways to treat the condition $b(t)=\frac{1}{\sqrt{\grL}}$: either as a constraint by introducing a Lagrange multiplier or by simply replacing it. The latter immediately yields
\begin{align}
L_{EH}|_{b(t)=\frac{1}{\sqrt{\grL}}}=0.\label{eq16}
\end{align}
Thus, the reduced Lagrangian is zero. Nevertheless, it is possible to construct a Lagrangian via which the equation \eqref{eq7} can be acquired. Since there is only one equation for two degrees of freedom $(n,a)$ or $(\dot{s},a)$ equivalently, we can choose only one of them as independent degree of freedom and the other as a function of the parameter $t$. So, instead of a singular Lagrangian, a time-dependent, regular one is obtained which reads
\begin{align}
L=\frac{\dot{a}^{2}}{2n(t)}+\frac{\grL}{2}n(t) a^{2},\label{eq17}
\end{align}
and mimics the Lagrangian of an inverse oscillator with ``mass''$=\frac{1}{n(t)}$ and a ``spring constant''$=\frac{\grL}{2}n(t)$. This Lagrangian is the unique, up to a multiplicative constant and the addition of a total derivative, quadratic in the ``velocity'' $\dot{a}$, whose Euler-Lagrange equations coincide with (10). Due to the covariance, the function $n(t)$ can be chosen at will so it is not an essential ``mass''.

Regarding the canonical Hamiltonian, we define the momentum as usual
\begin{align}
  p=\frac{\partial L}{\partial \dot{a}} = \frac{\dot{a}}{n}.\label{eq18}
\end{align}
Solving the previous relation with respect to $\dot{a}$, the Hamiltonian is obtained via the Legendre transformation
\begin{align}
&H=n(t)H_{0},\quad H_{0}=\frac{1}{2}\left(p^{2}-\grL a^{2}\right).\label{eq19}
\end{align}
While $H$ is not conserved due to the explicit time dependence of $n(t)$, $H_{0}$ is conserved. The Hamilton equations read
\begin{align}
\dot{a}=p\,n(t),\quad \dot{p}=\grL\,a\,n(t),\label{eq20}
\end{align}
and it can be proven that they are equivalent to the Euler-Lagrange equations.

\subsubsection{Case 2}

In this case, we can construct a reduced Einstein-Hilbert Lagrangian which is not zero. By substituting in the initial Lagrangian \eqref{eq6} the solution of the constraint \eqref{eq9}, the former acquires the square root form:
\begin{align}
L_{EH}|_{r}=2\sqrt{\grL b(t)^{2}-1}\sqrt{\dot{b}(t)\dot{w}}.\label{eq21}
\end{align}
It is straightforward to check that the two Lagrangians are equivalent.
The equation of motion \eqref{eq13} can be obtained via Euler-Lagrange equations either with respect to $w$ while $b(t)$ is considered a time-dependent function or the other way around. Since we want to quantize the system, using the canonical approach, it is to our interest to avoid a Lagrangian which depends on the square root of the velocity. It is not difficult,  through fitting some initially arbitrary functions of $w$ and $b$, to obtain the unique Lagrangian which is quadratic in the ``velocity'' $\dot{w}$ and yields the same Euler-Lagrange equation
\begin{align}
L_{s}=\frac{{\dot{w}}^{2}}{2(\grL\,b^{2}-1)\dot{b}}.\label{eq22}
\end{align}
We need to note that the classical equivalence between two Lagrangians, like \eqref{eq21} and \eqref{eq22}, which lead to the same Euler - Lagrange equations, does not necessarily guarantee a quantum equivalence as well. This is a valid problem recognized by several authors \cite{Lagequiv1,Lagequiv2}. It has also been noted that the process of quantization may be connected to the type of Lagrangian or Hamiltonian that is being used \cite{Lagequiv3}. For example the canonical quantization has been developed by taking in consideration Hamiltonians which are in general quadratic in the momenta and thus may fail for Hamiltonians with a different functional dependence. This is the main reason, why we choose to make the passing from Lagrangian \eqref{eq21} to \eqref{eq22}. In this manner we are going to have a quadratic Hamiltonian at our disposal. Thus, we may follow the typical process of canonical quantization avoiding the complex problem of assigning an appropriate operator to a Hamiltonian that has some different type of dependence in the momentum.

It is also interesting to mention that the question of quantum equivalence even plagues the process of the classical reduction from the original gravitational action to the minisuperspace description. In other words, it is not guaranteed that the minisuperspace Lagrangians, which correctly reproduce the Einstein's equations after adopting an ansatz for the line element, have a quantum description which is relevant to the quantum behaviour of the original gravitational system. The equivalence however of the classical dynamics, leads us to the assumption that, even at the worst case scenario, at least some properties of the quantum minisuperspace system may be of relevance and help us to obtain insights with respect to the actual quantum gravitational configuration. Besides, this is the main approach we have at our disposal in the absence of a complete theory of quantum gravity.

The expression \eqref{eq22} can be further simplified  by introducing another time dependent variable which transforms as a density under time reparametrizations, we thus define $s(t)$ as

\begin{equation} \label{defs}
  s(t)=\left(\grL b^{2}-1\right)\dot{b} \Rightarrow \int\!\! s(t) dt = \frac{\Lambda}{3} b^3 - b + \kappa,
\end{equation}
where $\kappa$ is a random integration constant. The Lagrangian is now written as
\begin{equation}
L_{s}=\frac{\dot{w}^{2}}{2s(t)},\label{eq223}
\end{equation}
and it is valid  for $b(t)\neq$constant. By the definition of $s(t)$, the forbidden value $b(t)=\frac{1}{\sqrt{\grL}}$ corresponds to $s(t)=0$.

The horizons can be found from the positive roots of the equation $w(t)=0$, where $w(t)$ is given by \eqref{eq14}. The relevant third order algebraic equation to be solved is
\begin{equation} \label{cubic}
  w=\frac{\Lambda}{3} b^3 - b + M =0 \Rightarrow b^3 - \frac{3}{\Lambda}b + \frac{3M}{\Lambda} =0.
\end{equation}
From the theory of cubic equations \cite{McKelvey}, we know that \eqref{cubic} can have up to three real roots. In particular, if $\Lambda<0$ there exists a single real root given by
\begin{equation}
  b_0 = -\frac{2}{\sqrt{-\Lambda}}\sinh \left[\frac{1}{3}\sinh^{-1}\left(\frac{3 M \sqrt{-\Lambda}}{2}\right) \right] ,
\end{equation}
which is positive when $M>0$.
In the case where $\Lambda>0$, there exists a single real root when $9\Lambda M^2- 4>0$ given by
\begin{equation}
  b_0 = -\mathrm{sign}(M) \frac{2 }{\sqrt{\Lambda}} \cosh\left[\frac{1}{3}\cosh^{-1}\left(\frac{3 |M| \sqrt{\Lambda}}{2}\right) \right] .
\end{equation}
The latter is positive when $M<0$. Finally, if $\Lambda>0$ and at the same time $9\Lambda M^2- 4<0$ there exist three real roots
\begin{align}
  b_0 & = - \frac{2}{\sqrt{\Lambda}} \cos\left[\frac{1}{3}\cos^{-1}\left(\frac{3 M \sqrt{\Lambda}}{2}\right)\right], \\
  b_{\pm} & =  \frac{1}{\sqrt{\Lambda}} \left\{ \cos\left[\frac{1}{3}\cos^{-1}\left(\frac{3 M \sqrt{\Lambda}}{2}\right)\right] \pm \sqrt{3} \sin\left[\frac{1}{3}\cos^{-1}\left(\frac{3 M \sqrt{\Lambda}}{2}\right)\right] \right\} .
\end{align}
The root $b_0$ is always negative, the roots $b_{\pm}$ are both positive when $M>0$, while for $M<0$ only $b_+$ is positive.

Behind the horizons the curvature singularity appears at $b=0$.
Regarding the Hamiltonian, it is simpler from what we had in the previous case
\begin{align}
&H=s(t)H_{0},\quad H_{0}=\frac{p^{2}}{2},\label{eq27}\\
&\dot{w}=p\,s(t),\quad \dot{p}=0.\label{eq28}
\end{align}
and, as we see, we have two conserved quantities, $H_{0}$ and $p$.

\section{Quantum description}

\subsection{Time-Covariant Schr\"{o}dinger equation}

The canonical quantization procedure will be followed in this section: the canonical momenta and positions are replaced by operators $(q^{i}\rightarrow\hat{q}^{i},\,p_{i}\rightarrow\hat{p}_{i})$ which satisfy the property of self-adjointness (given appropriate boundary conditions for the wave function),
\begin{align}
\left<\hat{p}_{j}\grc\lvert\grf\right>=\left<\grc\lvert\hat{p}_{j}^{*}\grf\right>,\,
\left<\hat{q}^{j}\grc\lvert\grf\right>=\left<\grc\lvert(\hat{q}^{j})^{*}\grf\right>,\label{eq29}
\end{align}
and the canonical commutation relations
\begin{align}
\left[\hat{q}^{j},\hat{q}^{l}\right]_{C}=0,\,
\left[\hat{q}^{j},\hat{p}_{l}\right]_{C}={\rm{i}}\, \grd^{j}_{l}\,,
\left[\hat{p}_{j},\hat{p}_{l}\right]_{C}=0,\label{eq30}
\end{align}
where $\left[\cdot,\cdot\right]_{C}$ denotes the commutator, $\grd^{j}_{l}$ the Kronecker's delta, $\left<\grc\lvert\grf\right>=\int{d^m q\,\grm \,\grc^{*}\,\grf}$ the inner product, where $m$ is the dimension of the configuration space and $\grm$ a proper measure. A usual choice for $\grm$ is to take the square root of the absolute value of the determinant of the configuration space metric (the so called natural measure). For the one dimensional Lagrangians \eqref{eq17} and \eqref{eq223}, which we derived earlier, the natural measure is a constant. Due to the invariance of the wave function up to a normalization constant, without loss of generality, we can simply take it as $\mu=1$. In the position representation of quantum mechanics the operators are chosen as
\begin{align}
&\hat{p}_{l}=-\rm{i} \partial_{q^{l}},\,\hat{q}^{l}=q^{l},\label{eq32}
\end{align}
where $q^{l}=(a,w)$.

As in the work \cite{Pailas:2020msz}, in both problems set by Lagrangians \eqref{eq17} and \eqref{eq223} we consider $t$ as an external time parameter and a Schr\"{o}dinger equation can be constructed which is time-covariant, with respect to time reparametrizations, due to the transformation properties of $n(t)$ and $s(t)$ respectively.

\subsubsection{Case 1}

For the moment, we will restore the standard S. I. base units to keep a better track of the scales involved. The Lagrangian of this case reads $L=\frac{\dot{a}^{2}}{2n}+\frac{\grL}{2}n a^{2}$. With our conventions, $a\sim{[\text{dimensionless}]},\,n\sim{[\text{dimensionless}]},\,\grL\sim{1/[\text{length}]^{2}}$ and so $L\sim{1/[\text{length}]^{2}}$ since $\dot{a}^{2}\sim{1/[\text{length}]^{2}}$. Note that, in our initial considerations, through which we derived $L$, we adopted the units $c=1$ and thus considered $x^0=t$. Here, in order to restore the S. I. units, we introduced again the time coordinate as, $t=x^{0}/c$, so the $\dot{a}$ in $L$ becomes $\dot{a}/c$. For the Lagrangian to acquire units of energy, we multiply it with $m c^{2}\grl^{2}$, where $m$ is some mass and $\grl$ some length unit. By also introducing the variable $q=\grl\,a$ we have the Lagrangian and Hamiltonian
\begin{align}
&\tilde{L}=\frac{m}{2n}\dot{q}^{2}+\frac{m\grv^{2}}{2}n\,q^{2},\label{eq333}\\
&\tilde{H}=n\left(\frac{p^{2}}{2m}-\frac{m\grv^{2}}{2}q^{2}\right),\label{eq5454}
\end{align}
where $\grv^{2}=\grL c^{2}$ and has units of frequency. The time-covariant Schr\"{o}dinger equation has the form
\begin{align}
{\rm{i}}\,\hbar \frac{\partial \grC(t,q)}{\partial t}=n(t)\left(-\frac{\hbar^{2}}{2m}\frac{\partial^{2}\grC(t,q)}{\partial q^{2}}-\frac{m\,\grv^{2}}{2}q^{2}\grC(t,q)\right).\label{eq775}
\end{align}
This is exactly the equation for an inverted harmonic oscillator of frequency $\grv={\rm{i}}\grv$. It has been extensively studied by previous authors, see for example \cite{Barton:1984ey},\cite{Guo_2011}. The general solution can be expressed in terms of the Parabolic Cylinder D functions,
\begin{align}
\grC(t,q)=e^{-{\rm{i}}E/\hbar \int{n(t)dt}}\left(\grm_{1}D[v,z]+\grm_{2}D[\bar{v},-\bar{z}]\right),\label{eq6567}
\end{align}
where $v=-\frac{1}{2}-\frac{{\rm{i}}E}{\hbar \grv},\,z=\left(1+{\rm{i}}\right)\sqrt{\frac{m\grv}{\hbar}}q$ and $\bar{v},\bar{z}$ their complex conjugates. If we turn to the classical solution \eqref{eq8}, we see that $a$ may attain - depending on the constants of integration - any value from $-\infty$ to $+\infty$. This is the domain of definition which we assign to $q$ as well. From a physical perspective though it can be argued, due to the metric being invariant under the change $a\mapsto -a$,  that this leads to calculating two times the same trajectories. If we restrict $a$ (and thus $q$) in the half-line $\mathbb{R}_+$, the differential operator $\frac{\partial^2}{\partial q^2}$ appearing in the Hamiltonian is not essentially self-adjoint. However, it admits a one-parameter family of self-adjoint extensions; the necessary conditions on the wave function are
\begin{equation}\label{boundaryhalfline}
  \Psi(t,0)-\nu \frac{\partial \Psi}{\partial q}\Big|_{q=0}=0,
\end{equation}
where $\nu$ is a constant parameter which models the behaviour at the boundary $q=0$ \cite{ClarkSharp,Fulop}. This obviously sets conditions over the integration constants $\mu_1$ and $\mu_2$. The well known Dirichlet and Neumann boundary conditions are obtained for $\nu=0$ and $\nu=\infty$ respectively. For the subsequent calculations that we make, which are centered on the propagator describing the time evolution of some initial state, the consideration of the half line - at least for an appropriate boundary condition - does not affect our result. We discuss this point later in Section \ref{sec41}. Thus, we consider the whole $\mathbb{R}$ as the domain of $a$ and $q$, so as to be in direct contact with the known results of the inverted oscillator and to cover all accepted solutions described by \eqref{eq8}; even if they are duplicate from the metric perspective, where just the $a^2$ appears.

Still, in the region $(-\infty,+\infty)$, the wave functions (41) are not square-integrable\footnote{Even though the wave functions are not square-integrable in $q \in (-\infty,+\infty)$ they are ``normalizable'' up to a delta function \cite{Barton:1984ey}.} and the energy spectrum of the Hamiltonian (39) is continuous, varying from minus to plus infinity. Due to the explicit dependence on the time appearing multiplicatively as a common factor in the Hamiltonian through $n(t)$, the former commutes with itself at different times. As a result, no time ordering needs to be considered for the time evolution operator. The constants $\grm_{1},\grm_{2}$ are subject to normalization, while the constant $E$ represents the ``energy'' of the particle. In the next section we will focus on some specific application for these wave functions.

Before proceeding we should comment on the arbitrariness of the time parameter appearing in \eqref{eq6567}. If we turn our attention to the action of the original minisuperspace system, eq. \eqref{eq5}, we know that its parametrization invariance implies a symmetry generator of the form $X = \chi\frac{\partial}{\partial t} - \dot{\chi} n \frac{\partial }{\partial n}$, with $\chi=\chi(t)$ an arbitrary function of time \cite{tchrismini}. The first part of the generator tells us that arbitrary  time transformations are in order, while the second encodes the information of the transformation law effected on the lapse by this reparametrization, which results in $n(t)dt$ transforming as a scalar. This symmetry is a remnant of the full diffeomorphism invariance of the full theory \cite{Ibragimov}. We observe that the wave function \eqref{eq6567} carries this remaining invariance of the minisuperspace system since its time dependence involves the integral of the factor $n(t)dt$. Of course it could be argued that one might reach the same result by applying an appropriate transformation over the standard time dependence, $e^{-\frac{{\rm{i}}E}{\hbar} t}$, of a typical Schr\"{o}dinger equation. However, that would be enforcing manually a symmetry, which we know that the original system has. In the process we follow here, this property is inherited naturally as a result of solving \eqref{eq775} without having chosen a particular time gauge. It is in this sense that we refer to an equation like \eqref{eq775} as a time-covariant Schr\"{o}dinger equation. As it is expected, the solutions and thus any kind of measurable quantity, for instance the probability current, is invariant under time reparametrizations and thus equivalent for different ``observers''.

\subsubsection{Case 2} \label{secwavecase2}

The units redefinitions are applied here as well. The Lagrangian for this case is $L=\frac{\dot{w}^{2}}{2s(t)}$. Due to the definitions so far, we have $w\sim[\text{length}],\,s\sim[\text{dimensionless}],\,L\sim[\text{dimensionless}]$. Restoring the usual time coordinate, multiplying with $mc^{2}$ and defining $q=w$ for the sake of simplicity, the Lagrangian and Hamiltonian read
\begin{align}
&\tilde{L}=\frac{m}{2s}\dot{q}^{2},\label{eq45556}\\
&\tilde{H}=s\frac{p^{2}}{2m}.\label{eq7558}
\end{align}
The corresponding time-covariant Schr\"{o}dinger equation becomes
\begin{align}
{\rm{i}}\,\hbar \frac{\partial \grC(t,q)}{\partial t}=n(t)\left(-\frac{\hbar^{2}}{2m}\frac{\partial^{2}\grC(t,q)}{\partial q^{2}}\right).\label{eq55543}
\end{align}
Note that from equation \eqref{eq12} it seems as if the variable $w$ has the same sign as $b$. Since we have assumed for simplicity (without loss of generality) $b>0$, this would imply that $w>0$. However, in the classical regime the solution $a^{2}<0$ is acceptable, thus even for $b>0$ we may have $w<0$. Therefore, there is no reason to restrict ourselves, meaning that $w$ is defined from minus to plus infinity. The situation here is quite different than what happens with the domain of $q$ in the previous section and the solution of Case 1. As we discussed previously, the $q<0$ of \eqref{eq333} is, from the gravitational perspective, duplicate to $q>0$. Here, the $q<0$ and $q>0$ (or $w<0$ and $w>0$ respectively) signify different regions of the same solution which are parted by horizons. Thus, we are dealing with distinct parts and not double copies of the same solution. Thus, we have no reservation to consider $q$ running in the whole real line.

Under this condition, equation \eqref{eq55543} expresses a free particle and the general solution can be written as
\begin{align}
\grC(t,q)=e^{-{\rm{i}}E/\hbar \int{n(t)dt}}\left(\grn_{1}\cos\left[\sqrt{\frac{2Em}{\hbar}}q\right]+\grn_{2}\sin\left[\sqrt{\frac{2Em}{\hbar}}q\right]\right),\label{eq3322}
\end{align}
where $\grn_{1},\grn_{2}$ constants subject to normalization and as before, $E$ is the ``energy'' of the particle.

\section{Invariant Probability of persistence}

We now attempt to make a deeper analysis on the wave functions obtained through the time-covariant Schr\"odinger equation. Our purpose is to study the evolution of initial states for the previously described systems and provide some interesting characteristics regarding the difference between the classical and quantum descriptions.

\subsection{Case 1} \label{sec41}

As it is evident from the previous sections, the equations describing the evolution of the scale factor coincide with those of the inverted harmonic oscillator. This characteristic evolution appears in cases when someone is interested to describe some unstable equilibrium. For instance, in our example, the scale factor follows an exponential dependence on time, since we have found that the line element reads
\begin{align}
ds^{2}=-n(t)c^2 dt^{2}+\left(c_{1}e^{\omega \int{n(t)dt}}+c_{2}e^{-\omega \int{n(t)dt}}\right)^{2}dr^{2}+\frac{1}{\grL}\left(d\gru^{2}+\sin^{2}\gru d\grf^{2}\right).\label{eq2893}
\end{align}
Thus, for $\int{n(t)dt}$ monotonically increasing, if $c_{1}=0$ the universe would grow exponentially, while for $c_{2}=0$ the universe would decrease exponentially and would require infinite time to reach a size of zero. On the other hand, the unstable equilibrium $a=0\Leftrightarrow c_{1}=0,c_{2}=0$ could persist for infinite time, which is classically acceptable as far as one stays in the inverted harmonic oscillator interpretation of the dynamics of the system. In what the geometry is concerned, the value $a=0$ renders the metric \eqref{eq2893} non invertible. All fourteen curvature scalars are polynomials of $\grL$ and thus constant. Space-times with this property are called Constant Scalar Invariants (CSI) spaces \cite{Coley}. We now proceed to study the problem and extract results based on the similarity with the inverted harmonic oscillator equation.

Quantum mechanically, the most we could do to describe the unstable equilibrium point, is to consider an initial state of minimum uncertainty localized around this point and study it's evolution, as well as how much time it is required for the universe to enter the exponential phase. To do so, we will follow \cite{Barton:1984ey} with some changes which will be discussed below.

The starting point is the initial state, which is taken to be a Gaussian, as well as the propagator of the inverted harmonic oscillator,
\begin{align}
&\grC(0,q)=\grp^{-1/4}l^{-1/2}e^{-q^{2}/(2l^{2})},\label{eq3322b}\\
&K(t,q,\tilde{q})=\left(\frac{m\grv}{2\grp {\rm{i}} \hbar \sinh(\grv B(t))}\right)^{1/2}\exp\left\{\frac{{\rm{i}}m \grv}{2\hbar \sinh(\grv B(t))}\left[\left(q^{2}+\tilde{q}^{2}\right)\cosh(\grv B(t))-2q\tilde{q}\right]\right\},\label{eq554}
\end{align}
where $l$ represents the width of the Gaussian wave-packet, $B(t)=\int{n(t)dt}$ and from now on the time dependence will be omitted. It can be verified that this propagator satisfies the equation \eqref{eq775} and the fact that for the initial state the uncertainty relation becomes minimum $\grD q \grD p=\frac{\hbar}{2}.$ Regarding the constant $l$, it is the standard deviation and some possible values will be discussed later on.

The evolution of the initial state is given by
\begin{align}
&\grC(t,q)=\int_{-\infty}^{+\infty}{d\tilde{q}K(t,q,\tilde{q})\grC(0,\tilde{q})},\label{eqqq33}\\
&\grC(t,q)=\grp^{-\frac{1}{4}}l^{-\frac{1}{2}}\left[\cosh(\grv B)+{\rm{i}}\left(\frac{r_{h}}{l}\right)^{2}\sinh(\grv B)\right]^{-1/2}\exp\left\{-\frac{1-{\rm{i}}2\gre^{2}\sinh(2\grv B)}{2 \left(r_{h}\grV\right)^{2}}q^{2}\right\},\label{eq3ree}
\end{align}
where the following quantities have been defined
\begin{align}
\gre^{2}=\frac{1}{4}\left[\left(\frac{r_{h}}{l}\right)^{2}+\left(\frac{l}{r_{h}}\right)^{2}\right],\,\grV^{2}=\left(\frac{r_{h}}{l}\right)^{2}\sinh^{2}(\grv B)+\left(\frac{l}{r_{h}}\right)^{2}\cosh^{2}(\grv B),\,r_{h}^{2}=\frac{\hbar}{m\grv}.\label{eq5ree}
\end{align}
The probability density and the uncertainty relation acquire a simple form
\begin{align}
&\grr :=\grC^{*}\grC=\grp^{-1/2}\left(r_{h}\grV\right)^{-1}\exp\left\{-\frac{q^{2}}{(r_{h}\grV)^{2}}\right\},\label{eqt55}\\
&\grD q\cdot \grD p=\frac{\hbar}{2}\sqrt{1+4\gre^{4}\sinh^{2}(2\grv B)}>\frac{\hbar}{2}.\label{ewdd2}
\end{align}
For the latter calculation we have used the well known relations for the root mean square deviations
\begin{equation}
  (\Delta q) = \left(\left< q^2 \right> - \left< q \right>^2 \right)^{\frac{1}{2}}, \quad (\Delta p) = \left(\left< p^2 \right> - \left< p \right>^2\right)^{\frac{1}{2}},
\end{equation}
in which the mean value for an observable $\widehat{A}$ is defined as  $\left< A \right> = \int \mu \Psi^*\widehat{A} \Psi dq$. Remember that in this case $\mu=1$.

Before proceeding, let us comment at this point that if we had adopted the half line as the domain of $a$ (and hence $q$). Then, we could use an image-like method \cite{Goodman,Dluhy} to write
\begin{equation}\label{halfprop}
  \tilde{K}(t,q,\tilde{q}) = K(t,q,\tilde{q}) + K(t,-q,\tilde{q}) .
\end{equation}
It is easy to see that such a propagator satisfies the Schr\"odinger equation \eqref{eq775} for a Neumann boundary condition $\partial_q\Psi(t,0)=0$, which corresponds to having $\nu=\infty$ in \eqref{boundaryhalfline}. Then, the integral \eqref{eqqq33} with $K(t,q,\tilde{q})$ in the real line $\mathbb{R}$ and the corresponding expression with $\tilde{K}(t,q,\tilde{q})$ in the half-line  $\mathbb{R}_+$ yield the same result for $\Psi$ as we see it in \eqref{eq3ree}. We need to recognize however that this correspondence appears with this specific ``reflective'' boundary condition. For different values of $\nu$ we expect different propagators and different results. Of course, even though we obtain the same $\Psi$ of \eqref{eq3ree} for this particular boundary condition, the integrals involving calculations that use this wave function have different domains of integration in the two cases. For the particular quantities regarding the time intervals which we calculate later, this just translates into a constant factor since the $\Psi$ of \eqref{eq3ree} is an even function in $q$. However, this factor is insignificant when considering the order of magnitude of these intervals as is going to be revealed later. It is thus our choice to remove altogether the ambiguity of $\nu$ in \eqref{boundaryhalfline} by considering the full mathematically accepted domain of definition for $a \in \mathbb{R}$, based on the classical solution.

\subsubsection{Invariant Non-decay probability and Mean lifetime.}

The probability that a system prepared at an initial state $\grC(0,q)$ has not yet decayed at the state $\grC(t,q)$ is calculated as follows
\begin{align}
&{\cal{P}}=|A|^{2},\,A=\int_{-\infty}^{+\infty}{dq\grC^{*}(0,q)\grC(t,q)},\label{eqyyh}\\
&A=\left[\cosh(\grv B)-\frac{{\rm{i}}}{2}\left[\left(\frac{l}{r_{h}}\right)^{2}-\left(\frac{r_{h}}{l}\right)^{2}\right]\sinh(\grv B)\right]^{-1/2},\label{eqddff}\\
&P=\left[1+4\gre^{4}\sinh^{2}(\grv B)\right]^{-1/2}.\label{eqwssx}
\end{align}
At this point, we define the \textit{\textbf{invariant}} mean lifetime as
\begin{align}
T_{M}:=\int_{0}^{\infty}{dB\,P}\Rightarrow T_{M}=\int_{0}^{\infty}{dB\,\left[1+4\gre^{4}\sinh^{2}(\grv B)\right]^{-1/2}},\label{eqdvbb}
\end{align}
or by introducing the variable $\grh=\grv B$
\begin{align}
T_{M}=\frac{1}{\grv}\int_{0}^{+\infty}{d\grh\,\left[1+4\gre^{4}\sinh^{2}(\grh)\right]^{-1/2}}.\label{esddd}
\end{align}

As it was argued in \cite{Barton:1984ey}, it is useful to give another realization of the decay rate via the following description: The probability Q, that a particle remains still within a distance $s_{p}$ is given by
\begin{equation}
\begin{split}
&Q(t,s_{p})=\int_{-s_{p}}^{+s_{p}}{dq |\grC(t,q)|^{2}}\Rightarrow \\
&Q(t,s_{p})=2\grp^{-1/2}\int_{0}^{\grj_{0}}{d\grj\,e^{-\grj^{2}}},\label{mmdd}
\end{split}
\end{equation}
where $\grj=\frac{q}{r_{h}\grV}$ and $\grj_{0}=\frac{s_{p}}{r_{h}\grV}$. In accordance to the previous case, we define an \textbf{\textit{invariant}} mean sojourn time as
\begin{align}
T_{Ms}=\int_{0}^{\infty}{dB\,Q(t,s_{p})}.\label{mskks}
\end{align}
It is rather helpful to define the following integration variable $\grz=\frac{l}{r_{h}\grV}$ such that $d\grz=-\grz\frac{1}{\grV}\frac{d \grV}{d B}dB$ and $\frac{1}{\grV}\frac{d \grV}{d B}=\grv \left(1-\grz^{2}\right)^{1/2}\left[1+\left(\frac{r_{h}}{l}\right)^{4}\grz^{2}\right]^{1/2}$ in order to have
\begin{align}
T_{Ms}=\frac{1}{\grv}\int_{0}^{1}{d\grz\frac{Erf\left(\frac{s_{p}}{l}\grz\right)}{\grz\sqrt{1-\grz^{2}}\sqrt{1+\left(\frac{r_{h}}{l}\right)^{4}\grz^{2}}}},\label{mkkff}
\end{align}
where $Erf\left(\frac{s_{p}}{l}\grz\right)$ the Error function defined as $Erf\left(\frac{s_{p}}{l}\grz\right)=2\grp^{-1/2}\int_{0}^{\frac{s_{p}}{l}\grz}{d \grj\, e^{-\grj^{2}}}$.

Both $T_{M}$ and $T_{Ms}$ depend on the ``frequency'' $\grv$ as we may have expected, since by definition $\grv$ is related to the only true length scale of the system $\frac{1}{\sqrt{\grL}}$. One of the differences is the introduction of an additional length scale $s_{p}$ in the second case, which might be related to the resolution distance of some apparatus.

\subsection{Case 2}

Let us now turn to the second case, where the Lagrangian resembles that of a free particle. The configuration for which $a=0\Leftrightarrow q=0$ corresponds to the horizon surface. Thus, we could follow the same procedure as previously, to find out how a configuration representing the horizon surface would evolve due to quantum mechanical effects. The same initial state is used as previously, while in this case, the propagator and the evolved state read
\begin{align}
&K(t,q,\tilde{q})=\left(\frac{m}{2\grp {\rm{i}}\hbar B(t)}\right)^{1/2}\exp\left\{\frac{{\rm{i}}m}{2\hbar B(t)}\left(q^{2}-\tilde{q}\right)^2\right\},\label{wsvvn}\\
&\grC(t,q)=\grp^{-\frac{1}{4}}l^{-\frac{1}{2}}\left(1+\rm{i}\grv_{p}B\right)^{-1/2}\exp\left\{-\frac{1-{\rm{i}}\grv_{p}B}{2l^{2}\grV^{2}}q^{2}\right\},\label{llmmdd}
\end{align}
where $\grv_{p}=\frac{\hbar}{m l^{2}},\,\grV^{2}=1+\left(\grv_{p}B\right)^{2}$ and $B(t)=\int{s(t)dt}$. The uncertainty relation becomes $\grD q \grD p=\frac{\hbar}{2}\grV>\frac{\hbar}{2}$.

\subsubsection{Invariant Non-decay probability and Mean lifetime}

Following the steps of the previous subsection, the desired quantities have the form
\begin{align}
&A=\left(1+\frac{{\rm{i}}}{2}\grv_{p}B\right)^{-1/2},\label{accv}\\
&{\cal{P}}=\left[1+\frac{1}{4}\left(\grv_{p}B\right)^{2}\right]^{-1/2},\label{fhjjj}
\end{align}
and so the mean life time becomes
\begin{align}
T_{M}=\frac{1}{\grv_{p}}\int_{0}^{\infty}{d\grh_{p}\,\left[1+\frac{1}{4}\grh_{p}^{2}\right]^{-1/2}}  = +\infty,\label{eqvvb}
\end{align}
where we used $\grh_{p}=\grv_{p}B$. The fact that the mean lifetime is infinite indicates that the initial Gaussian state is very stable. We comment on possible implications of this later in our analysis.

We proceed with the calculation of the mean sojourn time. By introducing the additional length scale, we acquire
\begin{align}
Q(t,s_{p})=2\grp^{-1/2}\int_{0}^{\grj_{0}}{d\grj\,e^{-\grj^{2}}} ,\label{eqsssx}
\end{align}
where $\xi=\frac{q}{l \omega}$ and $\grj_{0}=\frac{s_{p}}{l\grV}$. With the help of the variable $\grz_{p}=\frac{1}{\grV}$, the mean sojourn time reads
\begin{align}
T_{Ms}=\frac{1}{\grv_{p}}\int_{0}^{1}{d\grz_{p}\,\frac{Erf(\frac{s_{p}}{l}\grz_{p})}{\grz_{p}^{2}\sqrt{1-\grz_{p}^{2}}}}\label{effghh}.
\end{align}

\subsection{Specific values}

\subsubsection{Non-decay time and Mean sojourn time, Case 1}
In this subsection, some estimates will be provided for the non-decay time of the previously studied cases. Let us start from equation \eqref{esddd}
\begin{align}
T_{M}=\frac{1}{\grv}\int_{0}^{+\infty}{d\grh\,\left[1+4\gre^{4}\sinh^{2}(\grh)\right]^{-1/2}}.\label{dbggh}
\end{align}
Among the involved constants we have the mass $m$ which we introduced for dimensional reasons, the Planck constant $\hbar=1.1*10^{-34}\mathrm{J\cdot s}$, the speed of light $c=3*10^{8}\mathrm{m/s}$, the standard deviation $l$, the cosmological constant $\grL$ and the ``frequency'' $\grv^{2}=\grL\,c^{2}$.

At this point we assume the standard deviation of the original wavepacket to be of the order of the Planck length $l=l_{p}=1.6*10^{-35}\mathrm{m}$, where the quantum gravity effects are expected to appear. For the mass constant $m$ we choose a value which corresponds to the minimum energy content of the early universe $\sim 10^{19} \mathrm{GeV} = 1.6* 10^9 \mathrm{J} $. Thus, we use the mass we find by setting $mc^2 =1.6 * 10^9$, which yields $m= 1.8 * 10^{-8} \mathrm{kgr}$.

Regarding the cosmological constant, we will assume the most recent \cite{Aghanim:2018eyx} observed value $\grL\simeq7.5*10^{-52}\mathrm{m^{-2}}$. Based on these values, we acquire $\grv\simeq 8.2*10^{-18}\mathrm{s}^{-1}$, $r_{h}\simeq2.7*10^{-5}\mathrm{m}$ and $\gre\simeq8.5*10^{29}$. Due to the large value of $\gre^{4}$, the term $\gre^{4}\sinh^{2}(\grh)$ is almost zero even from the values $\grh\sim 10^{-52}$ and higher. With these values we calculate the integral of \eqref{dbggh} and the time $T_M$ to be
\begin{align}
&\int_{0}^{+\infty}{d\grh\,\left[1+4\gre^{4}\sinh^{2}(\grh)\right]^{-1/2}}\simeq4.6*10^{-59},\label{rgbv}\\
&T_{M}\simeq 5.6*10^{-42}\mathrm{s}.\label{rgbv2}
\end{align}
This is close to the order of magnitude of the Planck epoch $\sim 10^{-43}\mathrm{s}$. Of course this number refers to estimates based on a FLRW geometry, thus any comparison is to be taken with a precaution. We may assume that, the quantum packet was highly concentrated for times $t<T_{M}$ around the value zero, while for $t\geq T_{M}$ it starts decaying, leading at some point to an exponential dependence of the scale factor and thus driving some sort of inflationary epoch. How can we understand that the scale factor will acquire an exponential dependence? If the classical degree of freedom $q$ (or equivalently $a$) is connected to the mean value of the operator $\hat{q}$, then for the specific wave packet is equal to zero. However, we can obtain a non-zero expression based on the following estimate, $q_{class}^{2}=<{\hat{q}}^{2}>$
\begin{equation}
q_{class}^{2}=\frac{r_{h}^{2}\grV^{2}}{2}\Rightarrow q_{class} = \frac{r_h^2}{\sqrt{2} l} \sinh(\eta),\label{eqgbff}
\end{equation}
where we substituted the $\Omega$ as defined from \eqref{eq5ree}. This form of solution mimics the classical one \eqref{eq2893} for $c_{2}=-c_{1}$ and $c_{1}=\frac{r_{h}^{2}}{2\sqrt{2}l}$ (remember that $\eta=\omega B= \omega \int n(t)dt$). Thus, as time passes, the scale factor grows exponentially.

To see how other values affect the time $T_M$, let us take the value for the cosmological constant obtained via the standard model, $\grL\sim10^{70}\mathrm{m^{-2}}$, which is $10^{122}$ orders of magnitude larger. The time then is $T_{M}\sim 2.6*10^{-44}\mathrm{s}$, which is smaller than \eqref{rgbv2} and well inside the Planck epoch. If we keep $\grL$ as it is, and reduce the standard deviation, say take $l=9.7*10^{-2}l_{p}$, we obtain exactly the Planck time $T_{M}\simeq5.4*10^{-44}\mathrm{s}.$ Finally, with this $l$ and the cosmological constant from the standard model we obtain $T_{M}\simeq2*10^{-45}\mathrm{s}$. The important thing to notice is that all the relevant scales, lead to values of time in the domain where the quantum gravity effects are expected to be strong.

Regarding the mean sojourn time, it is given by the equation \eqref{mkkff}

\begin{align}
T_{Ms}=\frac{1}{\grv}\int_{0}^{1}{d\grz\frac{Erf\left(\frac{s_{p}}{l}\grz\right)}{\grz\sqrt{1-\grz^{2}}\sqrt{1+\left(\frac{r_{h}}{l}\right)^{4}\grz^{2}}}}.\label{afvggf}
\end{align}
An additional length scale was introduced, it is the length within which the wave packet is constrained, which we will assume to be some multiple $\grb$ of the Planck length. Furthermore, since the standard deviation is assumed to be equal to the Planck length, $\grb$ counts the number of times that standard deviation fits within the length:
\begin{align}
T_{Ms}=\frac{1}{\grv}\int_{0}^{1}{d\grz\frac{Erf\left(\grb\grz\right)}{\grz\sqrt{1-\grz^{2}}\sqrt{1+\left(\frac{r_{h}}{l_{p}}\right)^{4}\grz^{2}}}}
\end{align}
We provide the sojourn time for two values, $\grb=2,\grb=4$:
\begin{align}
&\grb=2\Rightarrow T_{Ms}\simeq1.4*10^{-41}\mathrm{s},\label{tgghh}\\
&\grb=4\Rightarrow T_{Ms}\simeq2.8*10^{-41}\mathrm{s}.\label{llokk}
\end{align}
As we can observe, the closer the value of $\grb$ to unity the closer the sojourn time to the non-decay time. The sojourn time $T_{Ms}$ for the inverted harmonic oscillator, and for a Gaussian wave packet, is decreasing for larger frequencies $\omega$. This is a known result in the theory of quantum mechanics. It is interesting to note however, that if the notion of  non-commutativity of space is introduced, then this changes and $T_{Ms}$ becomes a concave function of $\omega$, for further details on this see \cite{GuoRen}.

\subsubsection{Non-decay time and Mean sojourn time, Case 2}

For the case of the horizon decay, we follow pretty much the same procedure , with the same values. The parameter involved reads $\grv_{p}=4.3*10^{35}s^{-1}$. As we previously noted, the mean non-decay time here is infinite (see equation \eqref{eqvvb}). This seems to indicate that the initial state is extremely stable. We could infer that the previous case, which results in a $T_M$ in the range of \eqref{rgbv2}, is better suited for our purposes in the sense that it allows the universe to ``escape'' the initial state in a finite time period and to possibly transit to different phases of evolution. When comparing the two decay probabilities we observe that they have similar behaviours at initial times, i. e. $1-{\cal{P}} \propto t^2$ for both \eqref{eqwssx} and \eqref{fhjjj} with $B\propto t<<1$. This initial behaviour is referred to as a Zeno period in the literature \cite{Giacosa} as such time dependence leads to the known Zeno effect of Quantum Mechanics \cite{Misra}. At late times the two systems differentiate, the non-decay probability of equation \eqref{eqwssx} falls exponentially as $t\rightarrow +\infty$ while the one given by \eqref{fhjjj} diminishes as $t^{-1}$.

In contrast to the mean non-decay time, the sojourn time results in finite values of the same order to the previous case. We provide the result for two values of $\grb$
\begin{align}
&\grb=2\Rightarrow T_{Ms}=3.4*10^{-41} \mathrm{s},\label{tthdd}\\
&\grb=4\Rightarrow T_{Ms}=5.1*10^{-41} \mathrm{s}.\label{thjjh}
\end{align}
We see that this calculation yields the same order of values for $T_{Ms}$ as that of the previous case. We thus get a completely different picture by looking at this quantity when comparing to the non-decay time. It has been argued in \cite{Barton:1984ey}, for the system studied there, that the mean sojourn time can be considered more realistic than the non-decay time since the former takes into account an external length scale in its calculation (namely the $s_p$ of \eqref{mmdd}).

\section{The minisuperspace canonical quantization}

In this section, we briefly derive the wave functions that are obtained through the usual canonical quantization of the minisuperspace Lagrangian $L_{EH}$ of \eqref{eq6} by solving the relevant Wheeler-DeWitt equation. To a large extent the main formalism follows the procedure used in \cite{Schtchris} for the canonical quantum description of the Schwarzschild space-time.

For reasons that will become obvious later in our analysis, and which have to do with the construction of a typical eigenvalue problem, we multiply the Lagrangian $L_{EH}$ with a constant $\mathcal{V}_0$ to which we assign the description of a finite spatial volume. This is owed to the fact that the original Einstein - Hilbert action contains an integral over the spatial region, e.g. see equation \eqref{fromEHtomin}. We may take the latter as a constant
\begin{equation}
 \mathcal{V}_0=\int \sin\theta dr d\theta d\phi,
\end{equation}
by considering an integration in a finite region of the spatial section. If we start from Lagrangian $\mathcal{V}_0 L_{EH}$ and perform a reparametrization of the lapse function $n(t) \mapsto e(t)$ as
\begin{equation} \label{ntoe}
  n = \frac{\mathcal{V}_0 e}{2 a \left(1-\Lambda  b^2\right)},
\end{equation}
we are led to a new Lagrangian
\begin{equation} \label{einLag}
  L_e = \frac{1}{2 e} G_{ij} \dot{q}^i \dot{q}^j - e\frac{\mathcal{V}_0^2}{2} .
\end{equation}
The latter is equivalent to the initial Lagrangian, since the change $n(t) \mapsto e(t)$ just represents a time reparametrization. The configuration space is two dimensional, spanned by the $q^i =(a,b)$, and the ensuing minisupermetic is
\begin{equation} \label{minimet}
  G_{ij} = \begin{pmatrix}
             0 & 4 a b \left(1- \Lambda b^2 \right) \\
             4 a b \left(1- \Lambda b^2 \right) & 4 a^2 \left(1 - \Lambda b^2  \right)
           \end{pmatrix} .
\end{equation}

Thus, the Lagrangian $L_e$ can be interpreted as describing the motion of a free particle of mass $\mathcal{V}_0$ in a spacetime characterized by the metric $G_{ij}$. The $e$ in this formalism is called the einbein field \cite{Brink:1977}. The resulting Hamiltonian constraint for the theory is given by
\begin{equation} \label{miniham}
  \mathcal{H} = G^{ij} p_i p_j + \mathcal{V}_0^2 \approx 0,
\end{equation}
where $``\approx"$ denotes a weak equality in Dirac's theory of constrained systems \cite{Sund}. The Wheeler-DeWitt equation emerges by enforcing the quantum version of the Hamiltonian constraint upon the wave function as $\widehat{\mathcal{H}}\Psi =0$. For the factor ordering of the kinetic term of \eqref{miniham}, the optimal choice in what regards invariance of the probability under transformations in the configuration space and time reparametrizations like \eqref{ntoe}, is the conformal Laplacian, which however in two dimensions is trivially equivalent to the usual Laplacian. We can thus write
\begin{equation} \label{Laplacian}
  \widehat{\mathcal{H}} = -\frac{1}{2\mu} \frac{\partial}{\partial q^i}\left( \mu G^{ij} \frac{\partial}{\partial q^j}\right) + \mathcal{V}_0^2 ,
\end{equation}
where $\mu(q)=\sqrt{-\mathrm{\det}(G_{ij})}$ is the natural measure that appears in the definition of the inner product
\begin{equation}
  \left<\Phi | \Psi \right>= \int \mu(q) \Phi^*(q) \Psi(q) d^2q .
\end{equation}

We notice that the metric $G_{ij}$, given by \eqref{minimet}, describes a flat two dimensional geometry of hyperbolic signature. The relevant canonical quantization in such a space has been used before in several problems in the literature \cite{EAquant,Gielen,DimAndrChiral}. As basic observables, together with the Hamiltonian constraint, we can either use the commuting momenta in Cartesian-like coordinates or the boost in the relevant plane (in a similar manner to what is done with the angular momentum quantization in spaces with Euclidean signature). Thus, the choice of the solutions to the Wheeler-DeWitt equation is made by the use of the linear in the momenta integrals of motion of the classical system as basic observables. In other words, we construct operators based on the classical integrals of motion and use them to define eigenvalue equations to be used in conjunction with the Hamiltonian constraint.
\begin{enumerate}
  \item \textbf{Quantum description with respect to the momenta.} For the shake of simplicity we work in light-cone coordinates. Under the transformation $(a,b)\mapsto (u,v)$ where
    \begin{equation} \label{utoa}
      u = 2 a^2 b, \quad  v = b(1-\frac{\Lambda}{3}b^2),
    \end{equation}
    the metric \eqref{minimet} becomes $G_{ij}dq^i dq^j=2 du dv$. The minisuperspace Lagrangian is written as
    \begin{equation}\label{Lagminiuv}
      L_e = \frac{1}{e}\dot{u}\dot{v} -  e \mathcal{V}_0^2,
    \end{equation}
     and the Hamiltonian constraint reads
    \begin{equation} \label{HamCart}
      \mathcal{H} = p_u p_v + \mathcal{V}_0^2 \approx 0,
    \end{equation}
    where $p_u$ and $p_v$ are the momenta in the null directions. The quantization in these variables is straightforward. We consider $\widehat{p}_u = - \rm{i}\hbar \frac{\partial}{\partial u}$, $\widehat{p}_v = - \rm{i}\hbar \frac{\partial}{\partial v}$, while the $\widehat{u}$, $\widehat{v}$ just act multiplicatively. The classical constants of motion $p_u$, $p_v$ imply the quantum eigenvalue equations
    \begin{equation}
      \widehat{p}_u \Psi = k_u \Psi, \quad     \widehat{p}_v \Psi = k_v \Psi,
    \end{equation}
    which result to the well known plane wave solution $\Psi = \frac{1}{2\pi \hbar} e^{\frac{\rm{i}}{\hbar }\left(k_u u + k_v v\right)}$, where we have directly put the needed multiplicative constant for normalization. The subsequent enforcement of the Hamiltonian constraint $\widehat{\mathcal{H}}\Psi =0$ yields the following relation between the constants $k_u$ and $k_v$
    \begin{equation} \label{conuv}
      2 k_u k_v + \mathcal{V}_0^2 =0,
    \end{equation}
    which are thus related to the spatial volume.

    \item \textbf{Quantum description with respect to the boost.} The third, linear in the momenta, classical integral of motion that the system \eqref{HamCart} possesses is the boost $L_0 = u p_u - v p_v$, whose quantum equivalent can be written as $\widehat{L}_0 = -{\rm{i}} \hbar \left(u \partial_u - v \partial_v\right)$. It is however simpler if we go over to coordinates where $\widehat{L}_0$ is brought into normal form. This happens in the new variables $(u,v)\mapsto (\rho,\varphi)$ where
        \begin{equation}
          u = \frac{1}{\sqrt{2}}\left( \rho \cosh \varphi + \rho \sinh \varphi\right), \quad v = \frac{1}{\sqrt{2}}\left( \rho \sinh \varphi - \rho \cosh \varphi\right).
        \end{equation}
        The minisuperspace metric becomes now $G_{ij}=\mathrm{diag}(-1,\rho^2)$ and the boost operator is transformed to $\widehat{L}_0= -{\rm{i}} \hbar \frac{\partial}{\partial \varphi} $. The relevant eigenvalue equation,
        \begin{equation}
          \widehat{L}_0 \Psi = \lambda \Psi,
        \end{equation}
        results into a wave function of the form $\Psi(\rho,\varphi) = \frac{1}{\sqrt{2\pi \hbar}} e^{\frac{\rm{i}}{\hbar} \lambda \varphi} \psi(\rho)$. The unspecified function $\psi(\rho)$ is found with the help of the Wheeler-DeWitt equation that, by the use of \eqref{Laplacian}, leads to
        \begin{equation} \label{Bessel}
           \frac{1}{\rho}\frac{d}{d \rho}\left( \rho  \frac{d}{d \rho} \psi(\rho)\right) + \left( \frac{\lambda^2}{\rho^2} + \mathcal{V}_0^2 \right)\psi(\rho) =0,
        \end{equation}
        which is the Bessel equation with generic solution
        \begin{equation}
           \psi (\rho) = C_1 J_{{\rm{i}} \lambda} (\mathcal{V}_0 \rho) + C_2 Y_{{\rm{i}} \lambda} (\mathcal{V}_0 \rho),
        \end{equation}
        with $J_\mu(z)$, $Y_{\mu}(z)$ denoting the Bessel functions of the first and second kind respectively.

        We need to be careful in choosing the function that solves \eqref{Bessel}. If we express how the new variables $(\rho,\varphi)$ are related to the original $(a,b)$ of the minisuperspace, we have:
        \begin{equation} \label{newtoold}
          \rho = \sqrt{2} ab \sqrt{\frac{\Lambda}{3} b^2-1}  , \quad \varphi = \ln \left( \frac{a}{\sqrt{\frac{\Lambda}{3} b^2-1}}\right) .
        \end{equation}
        In the region where $\Lambda b^2 - 3\geq 0$ our variables are real. In this case, as a solution to \eqref{Bessel} we can choose the function $\psi(\rho)=\psi_{\lambda,\mathcal{V}_0} = \frac{C}{\cosh(\lambda\pi)} \mathrm{Re}[J_{{\rm{i}} \lambda} (\mathcal{V}_0 \rho)]$, where $\mathrm{Re}$ denotes the real part of the expression in the brackets. This function was firstly introduced and studied in \cite{Dunster}. It is normalizable up to a delta function
        \begin{equation}
          \int_0^{\infty} \! \rho\, \psi_{\lambda,\mathcal{V}_0'}^* \psi_{\lambda,\mathcal{V}_0} d\rho \propto \delta(\mathcal{V}_0-\mathcal{V}_0'),
        \end{equation}
        given the constraint set by the Hermiticity of the Hamiltonian operator $\mathcal{V}_0/\mathcal{V}_0'=e^{2 k \pi}{\lambda}$, $k\in \mathbb{Z}$ \cite{DimAndrChiral}. A different linear combination however of the solution can be chosen that evades conditions on the ``eigenvalue'' $\mathcal{V}_0$ \cite{Gryb2019,Gielen}. Note that the weight $\rho$ inside the integral is what originates from the natural measure in these coordinates since $\mu = \sqrt{- \mathrm{\det}G_{ij}}=\rho$.

        On the other hand, if we consider the case $\Lambda b^2 - 3<0 $, then in the above description $\rho,\varphi$ become complex. For $\varphi$ this is not a problem since the second of \eqref{newtoold} is written as $\varphi = - \frac{{\rm{i}}\pi}{2}+ \ln\left(\frac{a}{\sqrt{1-\frac{\Lambda}{3} b^2}}\right)$. At the level of the wave function, whose dependence on $\varphi$ is $\Psi \propto e^{\frac{\rm{i}}{\hbar} \lambda \varphi}$, this results in an extra multiplication constant which can be absorbed through the process of normalization. Things however are different for the variable $\rho$ which becomes imaginary and now the previously used Bessel combination cannot be normalized even up to a Dirac delta function. This problem is bypassed by adopting a different linear combination of the solution to the Bessel equation.  By considering $\rho={\rm{i}} \tilde{\rho}$, where now $\tilde{\rho} \in \mathbb{R}$, we may write the solution with respect to the modified Bessel function of the second kind
        \begin{equation}
          \psi(\tilde{\rho}) = C K_{{\rm{i}} \lambda} (\mathcal{V}_0 \tilde{\rho}).
        \end{equation}
        As before the $C$ denotes the relative normalization constant. The $K_{{\rm{i}} \lambda} (\mathcal{V}_0 \tilde{\rho})$ is a real function and the inner product yields the integral \cite{Gradshteyn}
        \begin{equation}
          \int_0^{\infty} \! \tilde{\rho}\, K_{{\rm{i}} \lambda}(\mathcal{V}_0' \tilde{\rho}) K_{{\rm{i}} \lambda}(\mathcal{V}_0 \tilde{\rho}) d\tilde{u} = \frac{\pi \left(\mathcal{V}_0'\mathcal{V}_0\right)^{-{\rm{i}}\lambda}\left[(\mathcal{V}_0')^{{\rm{i}}\lambda}-\mathcal{V}_0^{{\rm{i}}\lambda}\right]}{2\sin({\rm{i}} \lambda \pi)\left[(\mathcal{V}_0')^2-\mathcal{V}_0^2\right]},
        \end{equation}
        which is finite in the limit $\mathcal{V}_0'\rightarrow \mathcal{V}_0$.

        The critical limit $\Lambda b^2 - 3 = 0 $, which separates the two regions leading to different wave functions, is the outermost horizon which appears in the classical solution when the constant of integration $M$ is zero in \eqref{eq15}.

\end{enumerate}

\section{Discussion}

This work was dedicated to the study of the classical and quantum minisuperspace description for the $\grL$-Kantowski-Sachs universe. In the classical regime, by following the method developed in \cite{Pailas:2020msz}, regular, albeit time dependent Lagrangians (instead of the original singular Lagrangians) were constructed, which encode the dynamics for both branches of the solution space: The Bertotti-Kasner and the generic solution (the cosmological equivalent of Schwarzschild-de Sitter space).

The regular nature of the Lagrangian and the appearance of the external time parameter $t$, enabled the construction of a time-covariant Schr\"odinger equation. Covariant in the sense that the gauge freedom $t=f(\tilde{t})$ was not ``broken'' and hence different ``observers'' will, by default provide equivalent quantum descriptions. The general solutions to these equations were obtained and expressed in terms of elementary functions. Moving a step further from the work \cite{Pailas:2020msz}, a time-invariant non-decay probability was provided, alongside with the non-decay and mean sojourn time. Contrary to the usual definitions of the above objects, which are invariant at most under translations $t=\tilde{t}+\gre$, in our case the freedom is rather generic $t=f(\tilde{t})$.

The non-decay and mean sojourn time mentioned in the previous paragraph, were calculated for the following schemes: Regarding the Bertotti-Kasner solution, an initial Gaussian wavepacket was assumed to be concentrated around the value $a=0$ (where $a$ is the scale factor) and its evolution was obtained. By assuming the width of the wave packet to be of the Planck order and the observed value for the cosmological constant, the decay time was calculated to be in the region of $10^{-42}\mathrm{s}$. This is at the end of the Planck epoch. Furthermore, by choosing the value of the cosmological constant derived from the Standard Model of Particles, the decay time reduces to almost $\sim10^{-44}\mathrm{s}$, which is still inside the region where quantum gravity effects are expected to be significant.

For the generic solution, a Gaussian wavepacket was assumed to be concentrated around the horizon value for the scale factor. The non-decay time becomes infinite while the relative sojourn time provides a completely different picture. The order of magnitude for the latter, which is expected to be better suited for physical interpretation, is of the same order as the one of the special solution at $\sim 10^{-41}\mathrm{s}$.

We additionally derived the solutions from the usual Wheeler-DeWitt quantization, which led us to the following setting:
\begin{align} \label{Psi1}
  \Psi(u,v) & = \frac{1}{2\pi \hbar} e^{\frac{{\rm{i}}}{\hbar }\left(k_u u + k_v v\right)}, \quad 2 k_u k_v + \mathcal{V}_0^2=0,  \\ \label{Psi2}
  \Psi(\rho,\varphi) & \propto e^{\frac{{\rm{i}}}{\hbar} \lambda \varphi} \mathrm{Re}[J_{{\rm{i}} \lambda} (\mathcal{V}_0 u)] , \quad \Lambda b^2 - 3\geq 0,\\ \label{Psi3}
  \Psi(\tilde{\rho},\varphi) & \propto e^{\frac{{\rm{i}}}{\hbar} \lambda \mathrm{Re}(\varphi)} K_{{\rm{i}} \lambda} (\mathcal{V}_0 \tilde{\rho}) , \quad \Lambda b^2 - 3< 0 .
\end{align}
The first wave function corresponds to having the momenta of the minisuperspace as quantum observables in Cartesian-like coordinates and as a solution it was given in a previous work by J. Louko and T. Vachaspati \cite{exte1}. The last two expressions refer to quantization with respect to the boost, given for different regions of the classical solution, so that the relative variables remain real through the transformations we have performed.

Interestingly enough, the wave function \eqref{Psi1} is comparable to \eqref{eq3322}. The latter is also a plane wave solution in the variable $q=w$ which was used in section \ref{secwavecase2}. This is revealed through the following considerations: Let us notice that the first of \eqref{utoa} becomes $u=2q$, if we introduce the relation $a=\sqrt{w}{b}=\sqrt{q}{b}$. At the same time, the classical solution in these light-cone coordinates is written, in a parametrization invariant way, as being linear in $\int e(t)dt$. We may thus re-write \eqref{Psi1} as
\begin{equation} \label{eq3322new}
  \Psi(u,v(t)) = \frac{1}{2\pi \hbar} e^{- \frac{\rm{i}}{\hbar }\frac{\mathcal{V}_0^2}{2 k_u} \int e(t) dt}e^{\frac{\rm{i}}{\hbar } 2 k_u q },
\end{equation}
where we have used the, compatible with the classical solution, expression $v= \int e(t) dt$, i.e. we effectively consider $v$ as the time parameter of the problem, and we additionally substituted $k_v$ from \eqref{conuv}. The wave function \eqref{eq3322new} is essentially the same as that we obtained through a different process in \eqref{eq3322}. Of course the $e$ is not the same as the $n$ appearing in \eqref{eq3322}, but neither is the $t$ variable that is being integrated since the two functions are related through a time reparametrization.  Given this parametrization invariance of the problem we are allowed to make this comparison and relate the two results at a semiclassical level (in the sense that we utilized information from the classical solution).

As far as \eqref{Psi2} and \eqref{Psi3} are concerned, which regard the quantization with respect to the boost symmetry of the problem, we do not see them being recovered with the previous method. However, the analysis performed has the big advantage of offering the possibility of a quantum description for the special solution of Bertotti and Kasner, for which we have no minisuperspace approximation through \eqref{einLag} since $G_{ij}=0$.


\section*{Acknowledgements}
N. D. acknowledges the support of the Fundamental Research Funds for the Central Universities, Sichuan University Full-time Postdoctoral Research and Development Fund No. 2021SCU12117.

\bibliographystyle{unsrt}

\begin{thebibliography}{10}

\bibitem{10.2307/100496}
P.~A.~M. Dirac.
\newblock Generalized hamiltonian dynamics.
\newblock {\em Proc. R. Soc. Lond. A}, 246(1246):326--332, 1958.

\bibitem{Dirac:1958sc}
P.~A.~M. Dirac.
\newblock The theory of gravitation in hamiltonian form.
\newblock {\em Proc. Roy. Soc. Lond. A}, 246(1246):333--343, 1958.

\bibitem{PhysRev.116.1322}
R.~Arnowitt, S.~Deser, and C.~W. Misner.
\newblock Dynamical structure and definition of energy in general relativity.
\newblock {\em Phys. Rev.}, 116:1322--1330, 1959.

\bibitem{PhysRev.116.1324}
R.~Arnowitt, S.~Deser, and C.~W. Misner.
\newblock Quantum theory of gravitation: General formalism and linearized
  theory.
\newblock {\em Phys. Rev.}, 113:745--750, 1959.

\bibitem{PhysRev.116.1325}
R.~Arnowitt, S.~Deser, and C.~W. Misner.
\newblock Canonical variables for general relativity.
\newblock {\em Phys. Rev.}, 117:1595--1602, 1960.

\bibitem{PhysRev.116.1326}
R.~Arnowitt, S.~Deser, and C.~W. Misner.
\newblock Consistency of the canonical reduction of general relativity.
\newblock {\em J. Math. Phys.}, 1:434--439, 1960.

\bibitem{PhysRev.116.1327}
R.~Arnowitt, S.~Deser, and C.~W. Misner.
\newblock Energy and criteria for radiation in general relativity.
\newblock {\em Phys. Rev.}, 118:1100--1104, 1960.

\bibitem{PhysRev.116.1328}
R.~Arnowitt, S.~Deser, and C.~W. Misner.
\newblock Finite self-energy of classical point particles.
\newblock {\em Phys. Rev. Lett.}, 4:375--377, 1960.

\bibitem{PhysRev.116.1329}
R.~Arnowitt, S.~Deser, and C.~W. Misner.
\newblock Coordinate invariance and energy expressions in general relativity.
\newblock {\em Phys. Rev.}, 122:997--1006, 1961.

\bibitem{PhysRev.116.13210}
R.~Arnowitt, S.~Deser, and C.~W. Misner.
\newblock Wave zone in general relativity.
\newblock {\em Phys. Rev.}, 121:1556--1566, 1961.

\bibitem{Wheeler:1964qna}
J.~A. Wheeler.
\newblock Geometrodynamics and the issue of final state.
\newblock In C. DeWitt and B. DeWitt, editors, {\em Les Houches Summer Shcool of Theoretical Physics: Relativity,
  Groups and Topology}, pages 317--522. New York, Gordon and Breach, 1964.

\bibitem{Wheeler:1988zr}
J.~A. Wheeler.
\newblock Superspace and the nature of quantum geometrodynamics.
\newblock In L.-Z. Fang and R.~Ruffini, editors, {\em Quantum cosmology},
  volume~3, pages 27--92. World Scientific, 1987.

\bibitem{DeWitt:1962cg}
B.~S. DeWitt.
\newblock The quantization of geometry.
\newblock In Witten L., editor, {\em Gravitation: An introduction to current
  research}, pages 266--381. John Wiley and Sons, New York, London, 1962.

\bibitem{PhysRev.160.1113}
B.~S. DeWitt.
\newblock Quantum theory of gravity. i. the canonical theory.
\newblock {\em Phys. Rev.}, 160:1113--1148, Aug 1967.

\bibitem{PhysRev.162.1195}
B.~S. DeWitt.
\newblock Quantum theory of gravity. ii. the manifestly covariant theory.
\newblock {\em Phys. Rev.}, 162:1195--1239, Oct 1967.

\bibitem{PhysRev.162.1239}
B.~S. DeWitt.
\newblock Quantum theory of gravity. iii. applications of the covariant theory.
\newblock {\em Phys. Rev.}, 162:1239--1256, Oct 1967.

\bibitem{Schwarz_2007}
J.~H. Schwarz.
\newblock String theory: Progress and problems.
\newblock {\em Progress of Theoretical Physics Supplement}, 170:214–226,
  2007.

\bibitem{Ib_ez_2000}
L.~E. Ib\'a{\~n}ez.
\newblock The second string (phenomenology) revolution.
\newblock {\em Class. and Quantum Grav.}, 17(5):1117–1128, Feb 2000.

\bibitem{green_schwarz_witten_2012}
M.~B. Green, J.~H. Schwarz, and E.~Witten.
\newblock {\em Superstring Theory: 25th Anniversary Edition}, volume~1 of {\em
  Cambridge Monographs on Mathematical Physics}.
\newblock Cambridge University Press, Cambridge, 2012.

\bibitem{DUFF_1996}
M.J. Duff.
\newblock M theory (the theory formerly known as strings).
\newblock {\em Int. J. Mod. Phys. A}, 11(32):5623–5641, Dec 1996.

\bibitem{rovelli_2004}
Carlo Rovelli.
\newblock {\em Quantum Gravity}.
\newblock Cambridge Monographs on Mathematical Physics. Cambridge University
  Press, 2004.

\bibitem{Thiemann_2007}
T.~Thiemann.
\newblock Loop quantum gravity: An inside view.
\newblock {\em Lecture Notes in Physics}, vol 721, page 185–263, 2007.

\bibitem{Rovelli_1998}
C.~Rovelli.
\newblock Loop quantum gravity.
\newblock {\em Living Reviews in Relativity}, 1:1, Jan 1998.

\bibitem{Loll_1998}
R.~Loll.
\newblock Discrete approaches to quantum gravity in four dimensions.
\newblock {\em Living Reviews in Relativity}, 1:13, Dec 1998.

\bibitem{Ashtekar2021}
A.~Ashtekar and E.~Bianchi.
\newblock A short review of loop quantum gravity.
\newblock {\em Rep. Prog. Phys.}, 84:042001, 2021.

\bibitem{PhysRevD.40.2598}
W.~G. Unruh and R.~M. Wald.
\newblock Time and the interpretation of canonical quantum gravity.
\newblock {\em Phys. Rev. D}, 40:2598--2614, Oct 1989.

\bibitem{Kuchar:1991qf}
K.V. Kuchar.
\newblock Time and interpretations of quantum gravity.
\newblock {\em Int. J. Mod. Phys. D}, 20:3--86, 2011.

\bibitem{Kiefer:2013jqa}
C.~Kiefer.
\newblock Conceptual problems in quantum gravity and quantum cosmology.
\newblock {\em ISRN Math. Phys.}, 2013:509316, 2013.

\bibitem{Isham:1992ms}
C.~J. Isham.
\newblock Canonical quantum gravity and the problem of time.
\newblock {\em NATO Sci. Ser. C}, 409:157--287, 1993.

\bibitem{Giulia}
F. Di Gioia, G. Maniccia, G. Montani and J. Niedda.
\newblock Non-Unitarity problem in quantum gravity corrections to quantum field theory with Born-Oppenheimer approximation.
\newblock {\em Phys. Rev. D}, 103:103511, 2021.

\bibitem{H_hn_2020}
P.~A. H\"ohn and A.~Vanrietvelde.
\newblock How to switch between relational quantum clocks.
\newblock {\em New Journal of Physics}, 22(12):123048, Dec 2020.

\bibitem{Hoehn:2019owq}
P.~A. H{\"o}hn, A.~R.~H. Smith, and M.~P.E. Lock.
\newblock {The trinity of relational quantum dynamics}.
\newblock {\em Phys. Rev. D}, 104:066001, 2021.

\bibitem{Gielen}
S.~Gielen and L.~Men\'endez-Pidal.
\newblock Singularity resolution depends on the clock.
\newblock {\em Class. Quantum Grav.}, 37:205018, 2020.

\bibitem{Ryan}
M.~P. Ryan and L.~C. Shepley.
\newblock {\em Homogeneous relativistic cosmologies}.
\newblock Princeton University Press, Princeton, New Jersey, 1975.

\bibitem{Schwarzschild:1916uq}
K.~Schwarzschild.
\newblock On the gravitational field of a mass point according to einstein's
  theory.
\newblock {\em Sitzungsber. Preuss. Akad. Wiss. Berlin (Math. Phys. )},
  1916:189--196, 1916.

\bibitem{1916AnP...355..106R}
H.~Reissner.
\newblock {\"U}ber die eigengravitation des elektrischen feldes nach der
  einsteinschen theorie.
\newblock {\em Annalen der Physik}, 355(9):106--120, January 1916.

\bibitem{Hawking1984}
S.~W. Hawking.
\newblock The quantum state of the universe.
\newblock {\em Nucl. Phys. B}, 239:257, 1984.

\bibitem{Page:1990mh}
D.~N. Page.
\newblock {Minisuperspaces with conformally and minimally coupled scalar
  fields}.
\newblock {\em J. Math. Phys.}, 32:3427--3438, 1991.

\bibitem{Ashtekar2006}
A.~Ashtekar, T.~Pawlowski, and P.~Singh.
\newblock Quantum nature of the big bang: An analytical and numerical
  investigation.
\newblock {\em Phys. Rev. D}, 73:124038, 2006.

\bibitem{PintoNeto2007}
F.~T. Falciano, N.~Pinto-Neto, and E.~S. Santini.
\newblock An inflationary non-singular quantum cosmological model.
\newblock {\em Phys. Rev. D}, 76:083521, 2007.

\bibitem{Kiefer2012}
C.~Kiefer and M.~Kraemer.
\newblock Quantum gravitational contributions to the cmb anisotropy spectrum.
\newblock {\em Phys. Rev. Lett.}, 108:021301, 2012.

\bibitem{Vakili2012}
B.~Vakili.
\newblock Scalar field quantum cosmology: A schr{\"o}dinger picture.
\newblock {\em Phys. Lett. B}, 718:34, 2012.

\bibitem{Kim2014}
S.~P. Kim.
\newblock Third quantization and quantum universes.
\newblock {\em Nuclear Physics B (Proc. Suppl.)}, 246-247:68, 2014.

\bibitem{Gryb2019}
S.~Gryb and K.~P.~Y. Th\'ebault.
\newblock Bouncing unitary cosmology i. mini-superspace general solution.
\newblock {\em Class. Quantum Grav.}, 36:035009, 2019.

\bibitem{Kenmoku:1998ax}
M.~Kenmoku, H.~Kubotani, E.~Takasugi, and Y.~Yamazaki.
\newblock {de Broglie-Bohm interpretation for wave function of
  Reissner-Nordstrom-de Sitter black hole}.
\newblock {\em Int. J. Mod. Phys. A}, 15:2059--2076, 2000.

\bibitem{Christodoulakis:1991jd}
T.~Christodoulakis and E.~Korfiatis.
\newblock {Quantum mechanics of the general spatially homogeneous geometry
  coupled to a scalar field}.
\newblock {\em J. Math. Phys.}, 33:2863--2876, 1992.

\bibitem{Christodoulakis:2001um}
T.~Christodoulakis, T.~Gakis, and G.~O. Papadopoulos.
\newblock {Conditional symmetries and the quantization of Bianchi type I vacuum
  cosmologies with and without cosmological constant}.
\newblock {\em Class. Quant. Grav.}, 19:1013--1026, 2002.

\bibitem{Christodoulakis:2013sya}
T.~Christodoulakis, N.~Dimakis, Petros~A. Terzis, B.~Vakili, E.~Melas, and Th.
  Grammenos.
\newblock {Minisuperspace canonical quantization of the Reissner-Nordström
  black hole via conditional symmetries}.
\newblock {\em Phys. Rev. D}, 89(4):044031, 2014.

\bibitem{Dimakis:2016mpg}
N.~Dimakis, Petros~A. Terzis, A.~Zampeli, and T.~Christodoulakis.
\newblock {Decoupling of the reparametrization degree of freedom and a
  generalized probability in quantum cosmology}.
\newblock {\em Phys. Rev. D}, 94(6):064013, 2016.

\bibitem{Karagiorgos:2018gkn}
A.~Karagiorgos, T.~Pailas, N.~Dimakis, G.~O. Papadopoulos, Petros~A. Terzis,
  and T.~Christodoulakis.
\newblock {Quantum cosmology of Bianchi VIII, IX LRS geometries}.
\newblock {\em JCAP}, 04:006, 2019.

\bibitem{Kan2021}
N.~Kan, T.~Aoyama, T.~Hasegawa, and K.~Shiraishi.
\newblock {Eisenhart lift for minisuperspace quantum cosmology}.
\newblock {\em Phys. Rev. D}, 104:086001, 2021.

\bibitem{Corda:2019vuk}
C.~Corda and F.~Feleppa.
\newblock The quantum black hole as a gravitational hydrogen atom.
\newblock {\em arXiv:1912.06478}, 12 2019.

\bibitem{Corda:2020fjz}
C.~Corda, F.~Feleppa, and F.~Tamburini.
\newblock On the quantization of the extremal reissner-nordstrom black hole.
\newblock {\em EPL}, 132:30001, 2020.

\bibitem{Davidson:2014tda}
A.~Davidson and B.~Yellin.
\newblock {Quantum black hole wave packet: Average area entropy and temperature
  dependent width}.
\newblock {\em Phys. Lett. B}, 736:267--271, 2014.

\bibitem{Davidson:2012dt}
A.~Davidson and B.~Yellin.
\newblock {Schwarzschild mass uncertainty}.
\newblock {\em Gen. Rel. Grav.}, 46:1662, 2014.

\bibitem{Marugan1997}
G.~A. Mena~Marug\'an.
\newblock {Canonical quantization of the Gowdy model}.
\newblock {\em Phys. Rev. D}, 56:908, 1997.

\bibitem{Hybrid2008}
M.~Mart\'in-Benito, L.~J. Garay, and G.~A. Mena~Marug\'an.
\newblock {Hybrid quantum Gowdy cosmology: Combining loop and Fock
  quantizations}.
\newblock {\em Phys. Rev. D}, 78:083516, 2008.

\bibitem{Szek_quant}
A.~Paliathanasis, A.~Zampeli, T.~Christodoulakis, and M.~T. Mustafa.
\newblock {Quantization of the Szekeres system}.
\newblock {\em Class. Quantum Grav.}, 35:125005, 2018.

\bibitem{Pal_Zamp_Sz}
A.~Zampeli and A.~Paliathanasis.
\newblock {Quantization of inhomogeneous spacetimes with cosmological constant
  term}.
\newblock {\em Class. Quantum Grav.}, 38:165012, 2021.

\bibitem{exte1}
J. Louko and T. Vachaspati.
\newblock {On the Vilenkin boundary condition proposal in anisotropic universes.}
\newblock {\em Phys. Lett. B.}, 223, 21, 1989.

\bibitem{exte2}
J. Halliwell and J. Louko.
\newblock {Steepest-descent contours in the path-integral approach to quantum cosmology. III. A general method with applications to anisotropic minisuperspace models.}
\newblock {\em  Phys. Rev. D.}, 42, 3997, 1990.

\bibitem{exte3}
D. Conradi.
\newblock {Quantum cosmology of Kantowski - Sachs-like models.}
\newblock {\em  Class. Quantum Grav.}, 12, 2423, 1995.

\bibitem{Shojai2008}
A.~Shojai and F.~Shojai.
\newblock {$f(R)$ Quantum Cosmology}.
\newblock {\em Gen. Rel. Grav.}, 40:1967, 2008.

\bibitem{PaliathanasisBD}
A.~Paliathanasis, M.~Tsamparlis, S.~Basilakos, and J.~D. Barrow.
\newblock Classical and quantum solutions in brans-dicke cosmology with a
  perfect fluid.
\newblock {\em Phys. Rev. D}, 93:043528, 2016.

\bibitem{Xu2016}
M.-X. Xu, T.~Harko, and S.-D. Liang.
\newblock {Quantum cosmology of $f(R,T)$ gravity}.
\newblock {\em Eur. Phys. J. C}, 76:1, 2016.

\bibitem{Darabi2019}
F.~Darabi and K.~Atazadeh.
\newblock {$f(T)$ quantum cosmology}.
\newblock {\em Phys. Rev. D}, 100:023546, 2019.

\bibitem{Capozzielloft}
F.~Bajardi and S.~Capozziello.
\newblock Noether symmetries and quantum cosmology in extended teleparallel
  gravity.
\newblock {\em Int. J. Geom. Meth. Mod. Phys.}, 18:2140002, 2021.

\bibitem{Paliathanasisftb}
A.~Paliathanasis.
\newblock {Minisuperspace Quantization of $f(T,B)$ Cosmology}.
\newblock {\em Universe}, 7:150, 2021.

\bibitem{pokorski_2000}
S.~Pokorski.
\newblock {\em Gauge field theories}.
\newblock Cambridge Monographs on Mathematical Physics. Cambridge University
  Press, 2nd edition, 2000.

\bibitem{weinberg_1995}
S.~Weinberg.
\newblock {\em The Quantum Theory of Fields}, volume~1.
\newblock Cambridge University Press, 1995.

\bibitem{Maggiore_845116}
M.~Maggiore.
\newblock {\em A modern introduction to quantum field theory}.
\newblock Oxford master series in statistical, computational, and theoretical
  physics. Oxford Univ., Oxford, 2005.

\bibitem{PhysRev.126.1864}
R.~F. Baierlein, D.~H. Sharp, and J.~A. Wheeler.
\newblock Three-dimensional geometry as carrier of information about time.
\newblock {\em Phys. Rev.}, 126:1864--1865, Jun 1962.

\bibitem{Kuchar:1971xm}
K.~Kuchar.
\newblock Canonical quantization of cylindrical gravitational waves.
\newblock {\em Phys. Rev. D}, 4:955--986, 1971.

\bibitem{Brown:1994py}
J.~D. Brown and K.~V. Kuchar.
\newblock Dust as a standard of space and time in canonical quantum gravity.
\newblock {\em Phys. Rev. D}, 51:5600--5629, 1995.

\bibitem{Alexander:2012tq}
S.~Alexander, M.~Bojowald, A.~Marciano, and D.~Simpson.
\newblock Electric time in quantum cosmology.
\newblock {\em Class. Quant. Grav.}, 30:155024, 2013.

\bibitem{Bojowald:2010xp}
M.~Bojowald, P.~A. Hoehn, and A.~Tsobanjan.
\newblock An effective approach to the problem of time.
\newblock {\em Class. Quant. Grav.}, 28:035006, 2011.

\bibitem{Rovelli:1989jn}
C.~Rovelli.
\newblock Time in quantum gravity: Physics beyond the schrodinger regime.
\newblock {\em Phys. Rev. D}, 43:442--456, 1991.

\bibitem{Kiefer:1988ud}
C.~Kiefer.
\newblock Continuous measurement of intrinsic time by fermions.
\newblock {\em Class. Quant. Grav.}, 6:561, 1989.

\bibitem{Barvinsky:1993jf}
A.~O. Barvinsky.
\newblock Unitarity approach to quantum cosmology.
\newblock {\em Phys. Rept.}, 230:237--367, 1993.

\bibitem{Pons:2009cz}
J.~M. Pons, D.~C. Salisbury, and K.~A. Sundermeyer.
\newblock Revisiting observables in generally covariant theories in the light
  of gauge fixing methods.
\newblock {\em Phys. Rev. D}, 80:084015, 2009.

\bibitem{exte4}
K. Schleich.
\newblock {Is reduced phase space quantisation equivalent to Dirac quantisation?}
\newblock {\em  Class. Quantum Grav.}, 7, 1529, 1990.

\bibitem{exte5}
G. Kunstatter.
\newblock {Dirac versus reduced quantization: a geometrical approach.}
\newblock {\em  Class. Quantum Grav.}, 9, 1469, 1992.


\bibitem{Pailas:2020msz}
Theodoros Pailas.
\newblock {\textquotedblleft{}Time\textquotedblright{}-Covariant Schr\"odinger
  equation and the canonical quantization of the
  Reissner\textendash{}Nordstr\"om black hole}.
\newblock {\em Quantum Rep.}, 2(3):414--441, 2020.

\bibitem{Davidson3}
A. Davidson and B. Yellin.
\newblock Restoring time dependence into quantum cosmology.
\newblock {\em Int. J. Mod. Phys. D}, 21(11):1242011, 2012.

\bibitem{Barbour_2002}
J.~Barbour, B.~Z. Foster, and N.~\'O. Murchadha.
\newblock Relativity without relativity.
\newblock {\em Classical and Quantum Gravity}, 19(12):3217–3248, May 2002.

\bibitem{1998PhLA..245..363R}
W.~Rindler.
\newblock {Birkhoff's theorem with {\ensuremath{\Lambda}}-term and
  Bertotti-Kasner space}.
\newblock {\em Physics Letters A}, 245(5):363--365, August 1998.

\bibitem{PhysRev.116.1331}
B.~Bertotti.
\newblock Uniform electromagnetic field in the theory of general relativity.
\newblock {\em Phys. Rev.}, 116:1331--1333, Dec 1959.

\bibitem{Kantowski:1966te}
R.~Kantowski and R.~K. Sachs.
\newblock {Some spatially homogeneous anisotropic relativistic cosmological
  models}.
\newblock {\em J. Math. Phys.}, 7:443, 1966.

\bibitem{Alonso_Serrano_2014}
A.~Alonso-Serrano, L.~J. Garay, and G.~A. Mena~Marug\'an.
\newblock Correlations across horizons in quantum cosmology.
\newblock {\em Phys. Rev. D}, 90(12):124074, Dec 2014.

\bibitem{Lagequiv1}
M. Henneaux.
\newblock Equations of motion, Commutation relations and ambiguities in the Lagrangian formalism.
\newblock {\em Annals Phys.}, 140:45, 1982.

\bibitem{Lagequiv2}
S. Hojman and L. C. Shepley.
\newblock Equivalent Lagrangian in classical field theory.
\newblock {\em Foundations of Physics}, 16:465, 1982.

\bibitem{Lagequiv3}
F. J. Kennedy Jr. and E. H. Kerner.
\newblock Note on the inequivalence of classical and quantum Hamiltonians.
\newblock {\em American Journal of Physics}, 33:463, 1965.

\bibitem{McKelvey}
J.~P. McKelvey.
\newblock Simple transcendental expressions for the roots of cubic equations.
\newblock {\em Am. J. Phys.}, 52:269, 1984.

\bibitem{Barton:1984ey}
G.~Barton.
\newblock {Quantum mechanics of the inverted oscillator potential}.
\newblock {\em Annals Phys.}, 166:322, 1986.

\bibitem{Guo_2011}
Guang-Jie Guo, Zhong-Zhou Ren, Guo-Xing Ju, and Xiao-Yong Guo.
\newblock Quantum tunneling effect of a time-dependent inverted harmonic
  oscillator.
\newblock {\em Journal of Physics A: Mathematical and Theoretical},
  44(18):185301, apr 2011.

\bibitem{ClarkSharp}
T. E. Clark, R. Menikoff and D. H. Sharp.
\newblock Quantum mechanics on the half-line using path integrals.
\newblock {\em Phys. Rev. D}, 22:3012, 1980.

\bibitem{Fulop}
T. F\"ul\"op and I. Tsutsui.
\newblock A free particle on a circle with point interaction.
\newblock {\em Phys. Lett. A}, 264:366, 2000.

\bibitem{tchrismini}
T. Christodoulakis, N. Dimakis and Petros A. Terzis.
\newblock {Lie point and variational symmetries in minisuperspace Einstein gravity}.
\newblock {\em J. Phys. A: Math. Theor.}, 47:095202, 2014.

\bibitem{Ibragimov}
N. H. Ibragimov
\newblock {\em Transformation Groups Applied to Mathematical Physics}.
\newblock D. Reidel Publishing Company, Dordrecht, Boston, Lancaster 1985

\bibitem{Coley}
A. Coley, S. Hervik, and N. Pelavas.
\newblock {On Spacetimes with Constant Scalar Invariants}.
\newblock {\em Class. Quant. Grav.}, 23:3053, 2006.

\bibitem{Goodman}
M. Goodman.
\newblock {Path integral solution to the infinite square well}.
\newblock {\em Am. J. Phys.}, 49:843, 1981.

\bibitem{Dluhy}
P. Dluhy and A. Gangopadhyaya.
\newblock {Sharp and Infinite Boundaries in the Path Integral Formalism}.
\newblock {\em arXiv:1112.3674 [quant-ph]}, 12 2011.


\bibitem{Aghanim:2018eyx}
N.~Aghanim et~al.
\newblock {Planck 2018 results. VI. Cosmological parameters}.
\newblock {\em Astron. Astrophys.}, 641:A6, 2020.

\bibitem{GuoRen}
G.-J. Guo, Z.-Z. Ren, G.-X. Ju and C.-Y. Long.
\newblock The sojourn time of the inverted harmonic oscillator on the noncommutative plane.
\newblock {\em J. Phys. A: Math. Theor.}, 44:425301, 2011.

\bibitem{Giacosa}
F. Giacosa, P. Ko\'scik and Tomasz Sowi\'nski.
\newblock Capturing non-exponential dynamics in the presence of two decay channels.
\newblock {\em Phys. Rev. A}, 102:022204, (2020).

\bibitem{Misra}
B. Misra and E. C. G. Sudarshan.
\newblock{The Zeno’s paradox in quantum theory}.
\newblock {\em J. Math. Phys.} 18:756, (1977).

\bibitem{Schtchris}
T.~Christodoulakis, N.~Dimakis, Petros~A. Terzis, G.~Doulis, Th. Grammenos,
  E.~Melas, and A.~Spanou.
\newblock Conditional symmetries and the canonical quantization of constrained
  minisuperspace actions: the schwarzschild case.
\newblock {\em J. Geom. Phys.}, 71:127, 2013.

\bibitem{Brink:1977}
L.~Brink, P.~Di~Vecchia, and P.~Howe.
\newblock {A Lagrangian formulation of the classical and quantum dynamics of
  spinning particles}.
\newblock {\em Nucl. Phys. B}, 118:76, 1977.

\bibitem{Sund}
K.~Sundermeyer.
\newblock {\em Constrained dynamics}.
\newblock Springer - Verlag, Berlin, Heidelberg, New York, 1982.

\bibitem{EAquant}
N.~Dimakis, T.~Pailas, A.~Paliathanasis, G.~Leon, Petros~A. Terzis, and
  T.~Christodoulakis.
\newblock {Quantization of Einstein-aether scalar field cosmology}.
\newblock {\em Eur. Phys. J. C}, 81:152, 2021.

\bibitem{DimAndrChiral}
N.~Dimakis and A.~Paliathanasis.
\newblock Crossing the phantom divide line as an effect of quantum transitions.
\newblock {\em Class. Quantum Grav.}, 38:075016, 2021.

\bibitem{Dunster}
T.~M. Dunster.
\newblock Bessel functions of purely imaginary order, with an application to
  second-order linear differential equations having a large parameter.
\newblock {\em SIAM J. Math. Anal.}, 21:995, 1990.

\bibitem{Gradshteyn}
I.~S. Gradshteyn and I.~M. Ryzhik.
\newblock {\em Table of Integrals, Series, and Products}.
\newblock Elsevier, Amsterdam, Boston, Heidelberg, 7th edition, 2007.



\end{thebibliography}

\end{document}